\documentclass[journal,twocolumn]{IEEEtran}

\usepackage{epsfig,makeidx,color}
\usepackage{amsmath,amssymb,amsthm,bbm}
\usepackage{cite,graphicx}
\usepackage{enumerate}
\usepackage{hyperref}

\def\cX{{\cal X}}

\def\cG{{\cal G}}

\def\rH{{\rm H}}

\def\rT{{\rm T}}

\def\uZ{{\mathbb Z}}

\def\uP{{\mathbb P}}

\def\uE{{\mathbb E}}

\def\indicator{\mathbbm{1}}

\newtheorem{myproperty}{\bf Property} 

\def\be{ \begin{equation} }
\def\ee{ \end{equation} }
\def\bea{ \begin{eqnarray} }
\def\eea{ \end{eqnarray} }
\def\bx{{\bf x}}
\def\by{{\bf y}}

\def\bs{{\bf s}}
\def\ba{{\bf a}}

\def\bn{{\bf n}}

\def\bA{{\bf A}}

\def\bI{{\bf I}}

\def\bR{{\bf R}}

\def\bX{{\bf X}}

\def\b0{{\bf 0}}

\def\cC{{\cal C}}

\def\cN{{\cal N}}

\ifCLASSOPTIONonecolumn
  \newcommand{\figwidth}{0.60\columnwidth}
  \newcommand{\sfigwidth}{0.40\columnwidth}
\else
  \newcommand{\figwidth}{0.85\columnwidth}
  \newcommand{\sfigwidth}{0.70\columnwidth}
\fi

\begin{document}

\title{Evolutionary Game for Hybrid Uplink NOMA with
Truncated Channel Inversion Power Control}

\author{Jinho Choi and Jun-Bae Seo\\
\thanks{J. Choi is with
the School of Information Technology,
Deakin University, Geelong, VIC 3220,
Australia
(e-mail: jinho.choi@deakin.edu.au).}
\thanks{J.-B. Seo is with the Department of
Electrical Engineering, Indian Institute of Technology Delhi,
New Delhi, 110016, India (e-mail: jbseo@iitd.ac.in).}}

\date{today}
\maketitle

\begin{abstract}
In this paper, we consider hybrid uplink non-orthogonal multiple access
(NOMA) that can support more users 
by exploiting the notion of power-domain NOMA.
In hybrid uplink NOMA, we do not consider
centralized power control as 
a base station (BS) needs instantaneous channel state information
(CSI) of all users which leads to a high signaling overhead.
Rather, each user is allowed to perform power control under fading
in accordance with a truncated channel inversion power control policy.
Due to the lack of coordination of centralized power control, 
users in the same resource block compete for access. To analyze users' behavior,
evolutionary game can be considered 
so that each user can choose transmission strategies
to maximize payoff in hybrid uplink NOMA with power control.
Evolutionarily stable strategy (ESS) is
characterized with fixed costs as well as costs that depend on
channel realizations, and it is also shown that 
hybrid uplink NOMA can provide a higher throughput
than orthogonal multiple access (OMA).
To update the state in evolutionary game for
hybrid uplink NOMA, the replicator dynamic equation
is considered with two possible implementation methods.
\end{abstract}

{\IEEEkeywords
NOMA; Uplink Power Control; Evolutionary Game; Fading Channels}

\ifCLASSOPTIONonecolumn
\baselineskip 26pt
\fi

\section{Introduction} \label{S:Intro}

Non-orthogonal multiple access (NOMA) has 
been extensively studied as an alternative 
to conventional orthogonal multiple access (OMA) 
\cite{Dai15} \cite{Ding_CM} \cite{Choi17_ISWCS} \cite{Dai18}.
NOMA can be used for both uplink and downlink 
in cellular systems. For downlink NOMA,
a base station (BS) uses superposition coding 
with careful power allocation
to transmit signals to multiple users.
At users, successive interference cancellation (SIC)
is employed to decode signals. 
This approach is called power-domain NOMA because 
users' signals are differentiated by different power levels.
In \cite{Saito13} \cite{Kim13}, power-domain NOMA
is studied for downlink with beamforming in cellular systems.
For downlink millimeter-wave systems,
NOMA can also be employed with beamforming
as in \cite{Sun18} \cite{Xiao18}.
In \cite{Choi14} and \cite{Shin17}, downlink NOMA is applied to multiple
cells with coordinated multipoint transmission and beamforming,
respectively. In \cite{Sun18a},
distributed analog beamforming is considered 
to support cell-edge users as well as users close to
BSs for network NOMA (with multiple cells).
Note that in \cite{Sun19}, uplink NOMA is considered in a multi-cell scenario.

In \cite{Imari14}, power-domain NOMA is studied for uplink with centralized
power allocation. To assign the power levels
(to users) for successful SIC at the BS,
the BS needs to know all users' 
instantaneous channel state information (CSI).
However, in practice, full instantaneous
CSI may not be available at the BS.
For example, only long-term fading coefficient 
(i.e., statistical CSI)
can be available at the BS.
Thus, if power allocation is carried out with 
statistical CSI,
successful SIC is not guaranteed and there are outage events
as in \cite{Zhang16} 
(which is also true for downlink NOMA as in \cite{Choi17_CSI}),
and power allocation can be carried out to minimize
the impact of outage events as in
\cite{Liu18}.

In \cite{ChoiJSAC} \cite{Choi18_L}, uplink NOMA is seen
as a random access scheme, which is called NOMA-ALOHA,
where outage events happen due to collision in the power domain.
To decide access probabilities to different power levels
in NOMA-ALOHA, the notion of game theory
\cite{Fudenberg} \cite{Maschler13} is adopted in \cite{Choi_18G}
\cite{Seo18}.
In \cite{Seo19}, an evolutionary game approach
to NOMA-ALOHA is studied, where a large number of users
can choose strategies with certain probabilities to maximize their payoff.
In \cite{Ding19}, a NOMA-assisted grant-free 
access scheme is studied, in which grant-free users
can co-exist with grant-based users for uplink transmissions.

In this paper, we study a hybrid uplink NOMA system with
a large number of orthogonal radio resource blocks.
In each radio resource block, there are two users
competing for access as random access.
Compared to conventional uplink 
(i.e., uplink of OMA) where only
one user is allocated per radio
resource block, the number of users becomes doubled.
The rationale of the proposed approach 
is to achieve the same or higher 
spectral efficiency (defined later) with some additional transmit 
power cost spent by users, while supporting more users. In the proposed scheme, when one user does not transmit signals due to severe fading
under power control,
another user assigned to the same resource block can access successful. If two users have two different power levels as well as zero power level
(which means no transmission) so that the signals transmitted
by two users with different power levels can be successfully
decoded by SIC.
We use the truncated channel inversion power control
policy \cite{Goldsmith97a} with two non-zero target
receive power levels
and employ the notion of evolutionary game to 
decide the thresholds for 
truncated channel inversion power control that maximize
the average payoff.

The main contributions of the paper are as follows:
\emph{i)} hybrid uplink NOMA is proposed that
can effectively support more users by exploiting fading
with the spectral efficiency
that is higher than or equal to that 
of conventional uplink (of OMA);
\emph{ii)} an evolutionary game formulation is studied
and its solution is characterized to decide
thresholds for 
truncated channel inversion power control that is used in
hybrid uplink NOMA.

The rest of the paper is organized as follows.
In Section~\ref{S:SM}, the system model
for hybrid uplink NOMA is presented.
We formulate an evolutionary game for hybrid uplink NOMA 
to decide thresholds for 
truncated channel inversion power control 
in Section~\ref{S:Ngame}.
In Section~\ref{S:Anal},
the evolutionary game for hybrid uplink NOMA 
is analyzed to characterize solutions under
different settings. We discuss other issues
including comparisons with other schemes
and implementations in Section~\ref{S:OI}. 
Simulation results are presented in Section~\ref{S:Sim}.
We conclude the paper with some remarks in Section~\ref{S:Conc}.

\subsubsection*{Notation}
Matrices and vectors are denoted by upper- and lower-case
boldface letters, respectively.
The superscript $\rT$ and $\rH$ denotes the transpose and Hermitian transpose of a vector or matrix, respectively.
For a matrix $\bX$, $[\bX]_{m,n}$ represents the $(m,n)$th element of it.
We also denote by $\uE[\cdot]$ and ${\rm Var}(\cdot)$
 the statistical expectation and variance, respectively, whereas
$\cC \cN(\ba, \bR)$
represents the distribution of
a circularly symmetric complex Gaussian
(CSCG) random vector with mean vector $\ba$ and
covariance matrix $\bR$.

\section{System Model} \label{S:SM}

In this section, we consider an uplink system
based on power-domain NOMA with 
multiple (orthogonal) radio resource blocks.
In general,
when power-domain NOMA is applied to uplink,
there is a dilemma in terms of signaling overhead and spectral efficiency.
If a BS knows its users' 
(instantaneous\footnote{In the paper, CSI means
instantaneous CSI unless it is stated otherwise.}) CSI, 
it can decide users' 
transmit powers (and inform to users) for successful SIC,
which leads to a high spectral 
efficiency \cite{Imari14}. However, a high signaling overhead 
is expected to make CSI available at the BS under fading channels.
On the other hand, if the transmit powers are decided by
users, although there is no signaling overhead 
(to make CSI available at the BS),
a poor spectral efficiency or throughput is expected due to
outage events \cite{Zhang16}.
To address this dilemma, 
we consider a hybrid uplink NOMA scheme,
where
the BS arbitrarily allocates a resource block to two users regardless
of their CSI (as the BS does not have CSI).
In each radio resource block,
two users independently perform power control
under fading. When the two users experience
independent fading, a statistical multiplexing with random access is expected such that one user does not transmit signals due to severe fading, another 
user can access the radio resource block.
With the notion of power-domain NOMA, we generalize it in
this section.

Suppose that there are a group of users for uplink transmissions
with $M$ orthogonal radio resource blocks of capacity $F$.
While conventional OMA can support $M$ users, 
power-domain
NOMA can support more users by allocating the same
resource block to multiple users \cite{Ding_CM}.
Note that although the number of users per radio resource block can be large,
in this paper, we only focus on the case that there are
two users per radio resource block due to the limitation
of transmit power.
As mentioned earlier, the BS 
arbitrarily or blindly allocates each resource block to two users,
denoted by users 1 and 2, without knowing their CSI.

It is assumed that each user knows
his or her own CSI, but not the other's. 
Let $h_k (t)$ denote that the channel coefficient between
user $k \in \{1,2\}$ and the BS at time slot $t$. 
Throughout the paper, we assume block-fading channels \cite{TseBook05},
where the channel coefficient remains unchanged within a slot
interval and randomly varies from a slot to another
(thus, $h_k (t)$ and $h_k (t+1)$ become independent).
For convenience, we omit the index for time slot  $t$,
unless it is necessary.
The instantaneous signal-to-noise ratio
(SNR) is defined 
as $\gamma_k = \frac{|h_k|^2}{N_0}$,
which is known at user $k$.
In time division duplexing (TDD) mode, 
the BS can broadcast a pilot signal prior to uplink
transmissions so that each user is able to estimate the channel
coefficient, $h_k$, thanks to the channel reciprocity\footnote{On the
other hand, if the BS needs to estimate all users' 
instantaneous CSI,
each user should transmit a pilot signal,
which results in a prohibitively
high signaling overhead for a large number of users.}.
For uplink transmissions,
the transmit power can be decided at each user based on the CSI
or $\gamma_k$. 
For the power control over fading channels,
we employ the truncated channel inversion power control
\cite{Goldsmith97a}.
In particular, we assume that 
a user transmits his signal if $\gamma_k \ge \tau$, 
where $\tau > 0$ is a threshold (for power control)
to be discussed later.
In addition, when $\gamma_k \ge \tau$, 
a user can set the transmit power to
either $\frac{\rho_2}{\gamma_k }$ or $\frac{\rho_1}{\gamma_k}$ 
(which also depends on the instantaneous SNR as will be explained
later)
for power-domain NOMA, where $\rho_1$ and $\rho_2$
are the pre-defined receive power levels with $\rho_1 > \rho_2$.
 
To decide $\rho_1$ and $\rho_2$,
suppose that one user, say user 1, chooses the high transmit power
and the other, say user 2, chooses the low transmit power.
The received signal at the BS becomes
\be
\by = h_1 \sqrt\frac{\rho_1}{\gamma_1} \bs_1 
+ h_2 \sqrt\frac{\rho_2}{\gamma_2} \bs_2 + \bn,
\ee
where $\bs_k$ represents the (coded) signal block from user $k$
with $\uE[\bs_k] = 0$ and $\uE[\bs_k \bs_k^\rH] = \bI$, 
and $\bn \sim \cC \cN(0, N_0 \bI)$ is the background noise.
For SIC, the strong signal, i.e.,
the signal from user 1,
$\bs_1$, is to be decoded first. Once it is decoded, it can be removed 
from the received signal, $\by$, using SIC. Then, 
the BS can decode the other signal,
i.e., the signal from user 2, $\bs_2$.
To allow successful SIC and decoding,
$\rho_1$ and $\rho_2$ need to satisfy the following constraints:
\begin{align}\label{eq:eq1a}
\frac{\rho_1}{\rho_2 + 1} \ge \Gamma \quad \mbox{and}\quad
 \rho_2  \ge \Gamma,
\end{align}
where $\Gamma$ represents the 
signal-to-interference-plus-noise ratio 
(SINR) threshold for successful decoding.
If a capacity achieving code is used,
it is necessary to satisfy $\log_2 (1+ \Gamma) \ge r_{\rm tx}$,
where $r_{\rm tx}$ represents the transmission rate.
However, if a non-capacity achieving code is employed,
$\Gamma$ depends on the modulation order, code rate, and so on
\cite{Choi_18C}.
If the minimum powers are assigned 
for \eqref{eq:eq1a}, we have
\be
 \rho_2  = \Gamma \quad \mbox{and} \quad 
 \rho_1  = \Gamma (1 + \Gamma),
	\label{EQ:rNrN}
\ee
which implies that $\rho_1$ in dB has to be at least two times higher than
$\rho_2$ in dB.
Consequently, in order to avoid high $\rho_1$, 
the SINR threshold, $\Gamma$, cannot be too high.
With a moderate value of $\Gamma$ (e.g., 10 dB),
for successful decoding, channel coding is required as 
in uplink NOMA \cite{Choi_18C}. For example,
if 16-quadrature amplitude modulation (QAM)
is used with $\Gamma = 10$ dB, a channel code with a code
rate less than $\frac{ \log_2 (1+ \Gamma)}{\log_2 16}
= 0.8649$ is to be used.

Since each user in a radio resource block 
can independently determine the transmit power 
due to independent fading, 
if two users are more likely 
to choose different power levels (including zero transmit power),  
one or two signals can be expected to be transmitted successfully,
which can lead to a high throughput thanks to power-domain NOMA.

Note that, however, if the two users may choose
the same receive power level, it results in unsuccessful
SIC and no one can successfully transmit their signals.
Thus, in each resource block, 
contention-based multiple access\footnote{Since 
contention-based multiple access is used 
in each (radio resource) block,
it is easy to increase
the number of users per block (i.e.,  
a generalization with
more than two users per block is straightforward).
However, the throughput may decrease with the number of users, because
the probability that more than one user 
has the same power level increases. Therefore, it might
be reasonable to consider two users per block
unless another multiple access scheme to support more users
can be used.}
is used,
while the BS strictly allocates two users per radio resource block.
From this, the resulting scheme becomes a hybrid scheme
(as a limited contention-based multiple access for two
users per radio resource block is used together with
a deterministic allocation of two users for every radio
resource block) and is referred to as the hybrid uplink NOMA scheme.

For comparison with OMA, we
define the efficiency of the system bandwidth as 
the number of users supported by a unit bandwidth 
multiplied by channel usage over time, whereas $F$ 
denotes the overall bandwidth as mentioned earlier. 
First, let us consider that 
the users have always a packet to transmit. 
If the efficiency of the system bandwidth for OMA is  
denoted  by $e_o$, we obtain it as 
$e_o= \frac{M}{F}$. With the proposed scheme, if two users 
transmit their signals at receive
power level $\rho_1$ or $\rho_2$ 
with probability $0.5$, 
we can find two cases of collisions out of four outcomes; that is,  one user for $\rho_1$ (or $\rho_2$) and the other for $\rho_1$ (or $\rho_2$),  i.e., $(\rho_1,\rho_1)$ and $(\rho_2,\rho_2)$; further, we can see two outcomes for success as $(\rho_1,\rho_2)$, and  $(\rho_2,\rho_1)$. Since there can be $50$\% collisions in this case, when $e_h$ denotes the efficiency of the system bandwidth for hybrid NOMA, we have $e_h=0.5\times 2M/C_t=M/F$, 
which is equal to $e_o$. Thus, hybrid NOMA can support additional $M$ users
at the expense of additional transmit power 
for $\Gamma^2$ in \eqref{EQ:rNrN} compared to OMA 
while two systems have the same efficiency of the system bandwidth. 
Secondly, let us consider that each user has a packet 
to transmit with probability $\alpha$. 
Then, we have $e_o= \frac{\alpha M}{F}$ for OMA. 
On the other hand, in this case, $e_h$ for hybrid NOMA is expressed as
$e_h=\left(2\alpha(1-\alpha) + 0.5\alpha^2 \right) \frac{2M}{F}$,
where the first term indicates that 
a user has a packet, while the other does not. 
The second term shows that both users have a packet to transmit. 
Consequently, if $\alpha<1$, it always holds that $e_h>e_o$,
which demonstrates the superiority
of hybrid NOMA to OMA.

\section{Evolutionary Game for Hybrid Uplink NOMA}	\label{S:Ngame}

In this section, we focus on a power control approach at users
based on evolutionary game for the hybrid uplink NOMA scheme.
In particular, multiple actions are considered 
with the truncated channel inversion power control
so that the power control at a user can be carried out by selecting
an action, and its average payoff is obtained for evolutionary game.

\subsection{Power Control for NOMA}

For power-domain NOMA, the truncated channel inversion
power control is modified and
the transmit power can be given by
\be
P_k (\gamma_k)
= \left\{
\begin{array}{ll}
\frac{\rho_1}{\gamma_k}, & \mbox{if $\gamma_k > \tau_{\rm pn}$} \cr
\frac{\rho_2}{\gamma_k}, & \mbox{if $\tau < \gamma_k \le \tau_{\rm pn}$} \cr
0, & \mbox{if $\gamma_k \le \tau$,} \cr
\end{array}
\right.
	\label{EQ:Pk}
\ee
where $\tau_{\rm pn} > \tau$.
Note that $\tau_{\rm pn}$ is another threshold to be determined.
The resulting power control scheme can be seen as 
a generalized
truncated channel inversion power control for NOMA.

Accordingly, we can have the strategy set of the three 
actions. 
Action 1 is the transmission of high power, i.e., 
$P_k = \frac{\rho_1}{\gamma_k}$;
action 2 is the transmission of low power, 
i.e., $P_k = \frac{\rho_2}{\gamma_k}$;
and action 3 is no transmission, i.e., $P_k = 0$.
It is noteworthy that since each user's action 
is decided by $\gamma_k$, which is a random variable,
a user's selection of strategy can be seen 
as random to the other user.

\subsection{A Formulation of Evolutionary Game}
	
Let 
$x_i$ represent the probability of action $i \in \{1,2,3\}$.
According to the power control in \eqref{EQ:Pk},
we have 
\be
x_i = \Pr(\gamma_k \in \cG_i),
	\label{EQ:pxi}
\ee
where 
$\cG_1 = \{\gamma_k \,:\, \gamma_k > \tau_{\rm pn})$,
$\cG_2 = \{\gamma_k \,:\, \tau < \gamma_k \le \tau_{\rm pn})$,
and $\cG_3 = \{\gamma_k \,:\, 0 < \gamma_k \le \tau)$.
The probabilities of actions are dependent on $\tau$ and $\tau_{\rm pn}$.
In addition, let the set of the probabilities over the actions be
$\cX = \{\bx\,:\, \sum_{i=1}^3 x_i = 1, \ x_i \ge 0\}$,
where $\bx = [x_1 \ x_2 \ x_3]^\rT$ is the probability
distribution over 3 actions (or pure strategies).
A distribution $\bx$ is also called the state or profile
of the population.

In this section, consider a symmetric game with
the same reward and cost for each user.
Thus, we only focus on the payoff of user 1.
Denote by $R$ the reward\footnote{If a capacity achieving code
is employed, the achievable rate becomes
$\log_2 (1+ \Gamma)$. Thus, if we set
$R \propto \log_2 (1+ \Gamma)$, the reward
becomes proportional to the achievable rate for successful decoding.
Note that since the power levels
are decided as in \eqref{EQ:rNrN}, with both actions 1 and 2,
we can have the same reward, which is proportional to
$\log_2(1+\Gamma)$.}
when user $k$ successfully transmits
its signal. 
In addition, let $C_i (\gamma_k)$ be the cost
of action $i$. 
If user 1 succeeds to transmit its signal
with action $i \in \{1,2\}$, the payoff becomes $R - C_i (\gamma_1)$. 
On the other hand, if action 3 is chosen,
the payoff becomes $-C_3$, which is seen as the regret cost.

Note that we also consider the
case that $C_i$ is a pre-defined constant 
for each $i$ with $C_1 > C_2$
(because the cost of high transmit power is higher than
that of low transmit power).

Let us consider 
the average payoff of user 1, when user 2
employs the state $\bx$.
The average payoff of user 1
with action $i \in \{1,2\}$ is given by
\begin{align}
u_1 (i, \bx) 
=  R \uE_\bx [\indicator({\rm succeed \ with \ action}\ i)]
- \bar C_i,
	\label{EQ:u_1}
\end{align}
where $\uE_\bx [\cdot]$ is the expectation
with respect to the distribution $\bx$
and $\indicator(\cdot)$ represents the indicator function.
Here, if the cost depends on the instantaneous SNR,
we have
\be
\bar C_i = \uE[C_i (\gamma_1) \,|\, \gamma_i \in \cG_i], \ i = 1,2.
\ee
Otherwise, $\bar C_i = C_i$.
In addition, we have
\be
u_1 (3, \bx) = - C_3.
\ee

\begin{myproperty}	\label{L:1}
Let $\bar \bx = [\bar x_1\ \bar x_2\ \bar x_3 ]^\rT \in \cX$.
Then, 
the average payoff of user 1 with a state 
(mixed strategy) $\bar \bx$
becomes
\begin{align}
u (\bar \bx, \bx) 
= \sum_{i=1}^3 \bar x_i u_1 (i, \bx) 
= \bar \bx^\rT \bA \bx,
	\label{EQ:uxx}
\end{align}
where
\be
\bA = \left[
\begin{array}{ccc}
-\bar C_1 & R - \bar C_1 & R- \bar C_1 \cr
R -\bar C_2 & - \bar C_2 & R- \bar C_2 \cr
-C_3 & -C_3 & - C_3 \cr
\end{array}
\right].
\ee
\end{myproperty}
\begin{IEEEproof}
See Appendix~\ref{A:1}.
\end{IEEEproof}

Note that if we add $C_3$ to all the average payoffs,
the resulting payoff with action $i = 3$ becomes 0. Thus,
in the rest of the paper,
we assume that $C_3 = 0$ without loss of generality.

In the context of evolutionary game \cite{Weibull95},
the total number of users, $2M$, becomes the size of the population.
Let $\bar \bx$ be the state of the mutant
and $\bx$ be the state of the population,
where $\bar \bx \ne \bx$.
In addition, denote by $\epsilon \in (0,1)$ the size of 
the subpopulation of mutants.
Then, 
$u(\bar \bx, \epsilon \bar \bx + (1-\epsilon) \bx)$ becomes the average
payoff of a mutant.
Furthermore,
if there exists $\epsilon_{\rm max} \in (0,1)$ such that
\be
u(\bx, \epsilon \bar \bx + (1-\epsilon) \bx) >
u(\bar \bx, \epsilon \bar \bx + (1-\epsilon) \bx), \epsilon \in (0,
\epsilon_{\rm max}),
	\label{EQ:uuu}
\ee
$\bx$ is an evolutionarily stable strategy\footnote{In
evolutionary game theory \cite{Weibull95}, an 
ESS is a robust strategy which if adopted 
by a population cannot be invaded by any competing alternative strategy.
According to \eqref{EQ:uuu}, it is a local optimal
(power control) strategy 
corresponding to a local maximum payoff.} 
(ESS).
In \eqref{EQ:uxx}, 
$u(\bar \bx, \bx)$ can also be seen as
the payoff of user 1 with 
the mixed strategy $\bar \bx$ when 
$\bx$ is the mixed strategy of user 2.
Consequently, we can consider a two-person game 
for each radio resource block.

For the two-person game for each radio resource block,
we can also characterize a mixed strategy 
Nash equilibrium
(NE) \cite{Fudenberg} \cite{Maschler13}.

\begin{myproperty}	\label{L:2}
If $\bx^*$ satisfies
\be
u(\bx^*, \bx^*) \ge u_1(i, \bx^*), \ \mbox{for all} \ i \in \{1,2,3\}.
	\label{EQ:NE}
\ee
$\bx^*$ is a mixed strategy NE.
If we have the strict equality in \eqref{EQ:NE}, then
$\bx^*$ is a strict mixed strategy NE.
\end{myproperty}
\begin{IEEEproof}
See Appendix~\ref{A:2}.
\end{IEEEproof}

\section{Analysis}	\label{S:Anal}

In this section, we find the solutions to the hybrid uplink NOMA game
in Section~\ref{S:Ngame} in different settings.

\subsection{With Fixed Costs}

In this subsection, we consider the case that the costs are independent
of the SNR and pre-decided.

An ESS is also a mixed strategy NE \cite{Weibull95},
while the converse does not hold unless the game is symmetric.
Fortunately, since the game for each radio resource block 
is symmetric, we can have an ESS by finding a mixed
strategy NE.

\begin{figure}[thb]
\begin{center}
\includegraphics[width=\sfigwidth]{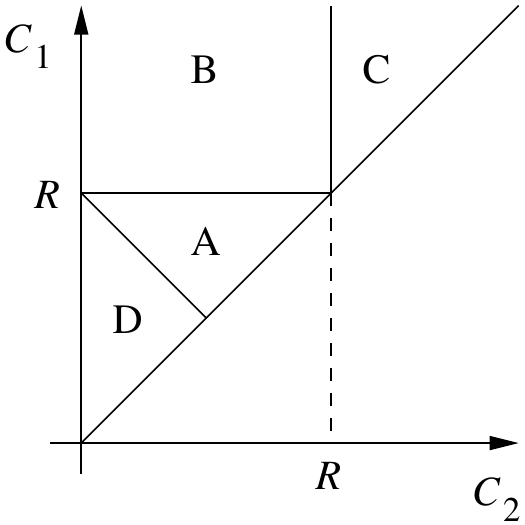}
\end{center}
\caption{Four solution regions for NE.}
        \label{Fig:regions}
\end{figure}

\begin{myproperty}	\label{L:3}
Suppose that Let $\Delta C = C_1 - C_2 > 0$.
There are 4 cases as follows:
\begin{itemize}
\item[A)] $C_1+ C_2 > R$ and $C_1 < R$: Then, we have
\be
(x_1^*, x_2^*, x_3^*) = \left(
1 - \frac{C_1}{R}, 1 - \frac{C_2}{R}, \frac{C_1+C_2}{R}  - 1\right).
	\label{EQ:L_A}
\ee
\item[B)] $C_1 > R$ and $0 < C_2 < R$: The solution is
\be
(x_1^*, x_2^*, x_3^*) = \left(0,
1 - \frac{C_2}{R}, \frac{C_2}{R} \right).
	\label{EQ:L_B}
\ee
\item[C)] $C_1, C_2 > R$: The solution is
\be
(x_1^*, x_2^*, x_3^*) = \left(0, 0, 1\right).
	\label{EQ:L_C}
\ee
\item[D)] $C_1+ C_2 < R$: The solution is
\be
(x_1^*, x_2^*, x_3^*) = \left(
\frac{1}{2} \left( 1 - \frac{\Delta C}{R}\right), 
\frac{1}{2} \left( 1 + \frac{\Delta C}{R}\right), 
0\right),
	\label{EQ:L_D}
\ee
where $x_1^* \in (0, \frac{1}{2})$
and $x_2^* \in (\frac{1}{2},1)$.
\end{itemize}
In Fig.~\ref{Fig:regions}, we show the 4 solution regions
(depending on the values of $C_1$ and $C_2$).
\end{myproperty}
\begin{IEEEproof}
See Appendix~\ref{A:3}.
\end{IEEEproof}

Once the $x_i$'s are obtained, we
can decide the values
for $\tau$ and $\tau_{\rm pn}$ using \eqref{EQ:pxi}. 
Thus, according to Property~\ref{L:3},
if $C_1+C_2 < R$ (i.e., region D),
$x_3^* = 0$ and $\tau = 0$, which means 
that a channel inversion power control 
without any truncation is to be used.
Clearly, this case is not desirable as the transmit power
can be very high for a small $|h_k|^2$ (or deep fading).
Furthermore, if $C_1 + C_2 > R$ (i.e., region C), no user
transmits signals as $x_3^* = 1$ and clearly this case should not be
considered. The case associated with region B 
is reduced to conventional truncated channel inversion power control
without power-domain NOMA, because only one target receive power level
(i.e., $\rho_2$) exists while $\tau_{\rm pn} \to \infty$.
Consequently, the case
associated with region A is suitable for hybrid uplink NOMA,  
where $x_1^*$ and $x_2^*$ decrease with their costs, $C_1$ 
and $C_2$, respectively.

As shown above, the ESS can be easily found
when the costs are fixed. However, 
since
we expect that the costs increase with the actual transmit power,
it might be more interesting to find
the ESS when costs are functions of the actual transmit power,
which is studied in the rest of the paper.

\subsection{With Costs Depending on Instantaneous SNR}
	\label{SS:C_SNR}

In this subsection, 
we study the case that the costs depends on the instantaneous SNR
under the assumption that the $h_k$'s are 
independent and identically distributed (iid) for tractable analysis.
In particular, we consider Rayleigh fading,
where 
\be
h_k \sim \cC \cN(0, \sigma_h^2), \ k \in \{1,2\}.
	\label{EQ:hk}
\ee

For the cost functions, we can consider the following ones:
\be
C_i (\gamma_k) = C \left(\frac{\rho_i}{\gamma_k}\right), \ i \in \{1,2\},
\ee
where $C(x)$ is an increasing function of $x$
so that the cost increases with the actual transmit power
in \eqref{EQ:Pk}.
In particular, if the energy efficiency is considered,
it is necessary to take into account the transmit power
for the cost so that the resulting strategy is more 
related to energy efficiency.
Closed-form expressions for the average cost functions 
are available
when $C(x)$ is a linear\footnote{Although
we only consider the case that the cost is a linear function
of the transmit power as in \eqref{EQ:Cx} in this paper,
it is also possible to consider another increasing function.
For example, if $C(p_{\rm tx}) = \ln (p_{\rm tx})$, 
where $p_{\rm tx}$ is the transmit power,
the payoff becomes
$R - \ln (p_{\rm tx}) = \ln \frac{e^R}{p_{\rm tx}}$. If 
$e^R$ is the transmission rate, the payoff becomes the energy 
efficiency in bits per second per transmit power. Thus,
the maximization of payoff is equivalent to the maximization
of energy efficiency.}
function as follows.

\begin{myproperty}	\label{L:4}
Suppose that
\be
C(x) = c x,
	\label{EQ:Cx}
\ee
where $c > 0$. For convenience, $c$ is referred to
as the scaling factor for costs.
Then, for the Rayleigh fading in \eqref{EQ:hk}, we have
\begin{align}
\bar C_1 & = \bar C_1 (\bx) =
\frac{c \rho_1}{\bar \gamma x_1}
E_1 \left( \ln \frac{1}{x_1} \right) \cr
\bar C_2 & = \bar C_2(\bx) =
\frac{c \rho_2}{\bar \gamma x_2}
\left( E_1 \left( \ln \frac{1}{x_1 + x_2} \right)
-E_1 \left( \ln \frac{1}{x_1} \right) \right), \quad 
	\label{EQ:CCs}
\end{align}
where $E_n (x) = \int_x^\infty \frac{e^{-z}}{z^n} dz$ is the exponential
integral and $\bar \gamma = \frac{\sigma_h^2}{N_0}$,
which is referred to as the average channel SNR.
\end{myproperty}
\begin{IEEEproof}
See Appendix~\ref{A:4}.
\end{IEEEproof}

Under Rayleigh fading, 
we need to have $\tau > 0$ in order to avoid infinite transmit power
\cite{Goldsmith97a}, which means that $x_3$
has to be greater than  $0$. This is also necessary
to avoid that $\bar C_2 (\bx)$
becomes infinite as shown in \eqref{EQ:CCs}. 
Therefore, when we consider Rayleigh fading
channels, it is desirable to have a non-zero $x_3$ or $\tau$.

\begin{myproperty}	\label{L:5}
Under a Rayleigh fading channel,
suppose that $x_3 > 0$.
Then, $x_1^*$ is the unique solution of
\be
R (1 - x_1) = 
\frac{c \rho_1}{\bar \gamma x_1}
E_1 \left( \ln \frac{1}{x_1} \right),
\ x_1 \in (0, \bar x_3),
	\label{EQ:cx1}
\ee
where $\tilde x_3 = 1 - x_3$, if the following condition holds
\be
\frac{c \rho_1}{\bar \gamma}
> \frac{R x_3 \tilde x_3}{ E_1 \left( \ln \frac{1}{\tilde x_3} \right)}.
	\label{EQ:x3_con}
\ee
In addition, $x_1^*$ decreases with $c$ and increases 
with $\bar \gamma$.
\end{myproperty}
\begin{IEEEproof}
See Appendix~\ref{A:5}.
\end{IEEEproof}

Once $x_1^*$ is found by solving \eqref{EQ:cx1},
$x_2^*$ can be found 
with known $x_1^*$. That is,
from \eqref{EQ:Rx23} and \eqref{EQ:CCs},
$x_2^*$ becomes the solution of the following equation:
\begin{align}
R(1- x_2) =
\frac{c \rho_2}{\bar \gamma x_2}
\left( E_1 \left( \ln \frac{1}{x_1^* + x_2} \right)
-E_1 \left( \ln \frac{1}{x_1^*} \right) \right), 
	\label{EQ:R2C2}
\end{align}
with $x_2 \in (0, 1- x_1^* - x_3)$.
Note that since $x_3$ is not known,
it is difficult to verify that the condition in
\eqref{EQ:x3_con} holds. Therefore, 
we can attempt to find $x_1^*$ by solving
\eqref{EQ:cx1} with $x_1 \in (0,1)$.
In this case, the solution always exists.
Then, we find $x_2^*$ by solving \eqref{EQ:R2C2}
with $x_2 \in (0, 1 - x_1^*)$. In this case,
as in Property~\ref{L:5}, we can show that $x_2^*$ exists and is unique.
If $x_1^* + x_2^* = 1$, the solution is not valid
(because $x_3$ becomes 0). In this case, a larger $c$ 
should be used to encourage non-transmission 
(i.e., $x_3 > 0$). 

Alternatively, in order to find the solution, we can use 
the replicator dynamic equation\footnote{In  
the replicator dynamic equation, 
$\dot{x}_i = \frac{d}{dt } x_i (t)$,
where $x_i (t)$ represents $x_i$ at time $t$. In 
a discrete-time system,
$\dot{x}_i = x_i (t + 1) - x_i (t)$,
where $t \in \uZ$ represents the discrete time unit.}
\cite{Weibull95} that is given by
\be
\dot{x}_i = \mu x_i ( u_1 (i, \bx) - u_1 (\bx, \bx)),
	\label{EQ:rde}
\ee
where $\mu > 0$ is the step-size.
We will consider 
the replicator dynamic equation from an implementation point of view
in Section~\ref{S:OI}.

\section{Other Issues}~\label{S:OI}

In this section, we consider
a few issues including comparisons with
other schemes that are not based on
game-theoretic setups and
a fairness issue in evolutionary game for hybrid uplink NOMA.

\subsection{Comparisons with Other Schemes}

The state or distribution, $\bx$, can be decided to maximize
the throughput that is the average number of successfully
transmitted users.
For a given $\bx$, the throughput 
of hybrid uplink NOMA per user can be found as
\begin{align}
\eta_{\rm hnoma} (\bx) 
& = 
\sum_{i=1}^2 \uE_\bx [\indicator(\mbox{succeed with action $i$})] x_i \cr
& = (x_2 + x_3) x_1 + (x_1 + x_3) x_2 \cr
& = (1-x_1) x_1 + (1-x_2) x_2,
	\label{EQ:eT}
\end{align}
which is a concave function of $x_1$ and $x_2$.
It can be readily shown that the following state maximizes
the throughput:
\be
x_1 = x_2 = \frac{1}{2}, \ x_3 = 0,
	\label{EQ:x2x3}
\ee
i.e., each user always transmits with action 1 or 2.
The maximum throughput per user becomes
\be
\eta^*_{\rm hnoma} = \max_\bx \eta_{\rm hnoma} (\bx) = \frac{1}{2}.
	\label{EQ:maxe}
\ee
and the total throughput (with two users) per resource block is 1.
Note that the total throughput of 1 
can also be achieved without power-domain NOMA, 
i.e., by allocating one user per radio resource block.
Therefore, hybrid uplink NOMA is not to increase
the throughput, but to support more users.
That is, the advantage of 
hybrid uplink NOMA over conventional uplink OMA 
is an increase in users to be supported (with the same throughput).

Note that
in practice, it is difficult to achieve the total throughput of 1 
with or without power-domain NOMA due to transmit power constraints
under fading.
To see this, consider \eqref{EQ:x2x3}, where $x_3 = 0$ or $x_1+x_2 = 1$.
Under the Rayleigh fading in \eqref{EQ:hk},
the average power
with action 2 becomes $\infty$ from \eqref{EQ:CCs}
(when $x_1+x_2 = 1$, $C_2 (\bx) = \infty$ since $\lim_{z \to 0} E_1(z) = 0$).
Therefore, it is necessary to keep $\tau > 0$ or $x_3 > 0$.
This is also true for the case without power-domain NOMA.
With a certain non-zero threshold $\tau > 0$
and its corresponding $x_3 = \delta > 0$,
we can consider
time division multiple access (TDMA) for two users per radio
resource block. In this case, 
the throughput per user becomes 
\be
\eta_{\rm tdma} = \frac{1 - \delta}{2}.
\ee
For hybrid uplink NOMA, 
from \eqref{EQ:eT}, after some manipulations,
we can have
\begin{align}
\eta_{\rm hnoma}^*  
& = \max_{x_1 + x_2 \le 1 - \delta} (1-x_1) x_1 + (1-x_2) x_2 \cr
& = \frac{1 - \delta^2}{2}.
\end{align}
This indicates that when truncated channel inversion power control 
is employed with a non-zero threshold, $\tau > 0$, 
hybrid uplink NOMA can provide a higher throughput than
OMA by a factor of 
up to
$\frac{1 - \delta^2}{1 - \delta} = 1+\delta
= 1+ x_3$.
Clearly, as in \eqref{EQ:tts}, $\tau$ increases with $x_3$
under Rayleigh fading. From this, if users have transmit power
constraints and need to keep a high threshold, $\tau$, 
hybrid uplink NOMA is preferable to OMA 
as it can effectively allow to share the radio resource block
between two users and improve the throughput.
Note that 
centralized power control, which requires
CSI from all the users, is not used in
both hybrid uplink NOMA and OMA. As a result,
hybrid uplink NOMA has signal overhead comparable to OMA,
while its throughput can be higher than that of OMA.

%

\subsection{Fairness in Evolutionary Game for Hybrid Uplink NOMA}
\label{SS:Fair}

In Subsection~\ref{SS:C_SNR}
we consider the case that
the cost functions depend on the instantaneous SNR.
As shown in \eqref{EQ:CCs},
the average cost, $\bar C_i,\ i \in \{1,2\}$, is
shown to be inversely proportional to 
the average channel SNR, $\bar \gamma$.
In general, the average channel SNR is decided
by the large-scale fading term
that is inversely proportional to the distance
between the BS and the user.
This implies that the cost of the user close to the BS (called
near users)
is smaller than that of user far away from the BS (called
far users).
Consequently, 
near users can take advantage of low costs
and will have higher transmission probabilities 
than far users.
Certainly, this results in unfairness in transmission
opportunities,
and fairness policies 
\cite{Timotheou15} \cite{Choi16_F}
are needed to be imposed.

In the evolutionary game for hybrid uplink NOMA,
we can impose the fairness 
by letting the value of $c$ 
in \eqref{EQ:Cx} be proportional to
the average channel SNR at each user,
i.e., $c \propto \bar \gamma_k$,
where $\bar \gamma_k$ represents the average
channel SNR at user $k$.
Then, in \eqref{EQ:CCs}, we can see that
$\bar C_1$ and $\bar C_2$ become independent of $\bar \gamma_k$
and a fairness can be achieved (i.e., the same state
at every user).

\subsection{Implementation of State Updating}	\label{SS:Impl}

If the BS knows statistical CSI of fading channels,
(e.g., the pdf of $|h_k|^2$ in \eqref{EQ:hk}),
it can decide the ESS,
$(x_1^*, x_2^*, x_3^*)$, by solving \eqref{EQ:cx1} and \eqref{EQ:R2C2}
with closed-form expressions for $u_1(i,\bx)$, $i \in \{1,2\}$,
and broadcasts it to all the users so that each user
can play the evolutionary game 
(or perform the modified
truncated channel inversion power control)
for hybrid uplink NOMA with the ESS. 
However, in practice,
it may be difficult for the BS to 
have statistical CSI of fading channels (which may also slowly vary)
in advance.
Thus, closed-form expressions for $u_1(i,\bx)$
are not available. In this case, the BS is forced to use 
the replicator dynamic equation in \eqref{EQ:rde}
to find the ESS with estimates of $R_i (\bx)$ 
and $\bar C_i (\bx)$, $i\in \{1,2\}$. 
For a given $\bx$, at time slot $t$, we consider
the following estimate of $R_i (\bx)$:
\be
\hat R_i(t;\bx)  = \frac{R}{2M} \sum_{m=1}^M  Y_m(t;i),
	\label{EQ:hRi}
\ee
where $Y_m(t;i) \in \{0,1,2\}$ is the number of successfully
decoded signals in resource block $m$ at time slot $t$,
which is available at the BS. 
Since the average cost is available at users,
we assume that each user sends their average
costs for actions 1 and 2 once in a block consisting of $B$ slots,
where $B \ge 1$. 
Here, $B$ becomes the time window for time average.
If $B$ increases, the feedback rate 
to send the average cost decreases and leads to a lower\footnote{Compared
to centralized power control, where the CSI is to be updated
at every slot, the feedback rate becomes lower by a factor of $B$.} 
signaling overhead (at the cost of delayed state updates).
Together with the average costs from the users
and the estimate of $R_i (\bx)$ in
\eqref{EQ:hRi}, the estimates of $u_1 (i, \bx)$,
$i = 1,2$, become available at the BS.
Provided that the channels are iid and their statistics
are invariant over the time,
the BS can find the ESS after a number of iterations.
This approach is also valid 
even if each channel has a different
average channel SNR
as long as the scaling factor for costs is decided to be 
proportional to $\bar \gamma_k$, i.e.,
$c = c_k \propto \bar \gamma_k$ at each user, which 
results in the $\bar C_i$'s being
independent of the average channel SNR at each user
as mentioned in Subsection~\ref{SS:Fair}.
The resulting approach is referred to
as state updating at the BS (SU-BS).

To avoid the uplink signaling overhead to send 
the time averages of costs from users, 
we can further consider an approach that each user employs
the replicator dynamic equation to update
its own state to approach the ESS.
In this case, the BS needs to send acknowledgment (ACK)
or negative acknowledgment (NACK) signal
back to users at the end of each slot.
From the feedback information, two users 
are able to estimate
$R_i$ as in \eqref{EQ:hRi}.
Let $b$ denote the block index.
For each user in 
radio resource block $m$ (we now omit the index $m$),
we have the time average of payoff as follows:
\be
\bar u_k [i,\bx; b] = 
\frac{1}{B} \sum_{t=bB}^{(b+1)B-1} \left( \hat R_i (t; \bx)
- c_k \frac{\rho_i }{\gamma_k (t)} 
\right), \ i \in \{1,2\}.
	\label{EQ:TA_PO}
\ee
Then, using the time average of payoff in 
\eqref{EQ:TA_PO}, each user can update the
state at a block rate as follows:
\begin{align}
& x_{k,i} [b+1] - x_{k,i} [b] \cr
& = \mu x_{k,i} [b]
\left(
\bar u_k [i, \bx_k [b]; b]
 - 
\bar u_k [\bx_k[b], \bx_k [b]]
\right),
\end{align}
where $x_{k,i} [b]$ denotes
the probability of action $i$ at user $k$ in the $b$th block and
\be
\bar u_k [\bx_k[b], \bx_k [b]]
= \sum_{i=1}^3 x_{k,i} [b] 
\bar u_k [i, \bx_k [b]; b].
\ee
The resulting approach is referred to
as state updating at users (SU-U).

\section{Simulation Results and Discussions}	\label{S:Sim}

In this section, we present simulation results 
when the cost is
inversely proportional to the instantaneous SNR\footnote{Since the ESS 
is fully characterized 
in Property~\ref{L:3} when the costs are fixed,
we do not consider the case of fixed costs in this section.} 
as in \eqref{EQ:Cx}
under Rayleigh fading channels 
with iid $h_k$ in \eqref{EQ:hk} 
for all users.

\begin{figure}[thb]
\begin{center}
\includegraphics[width=\figwidth]{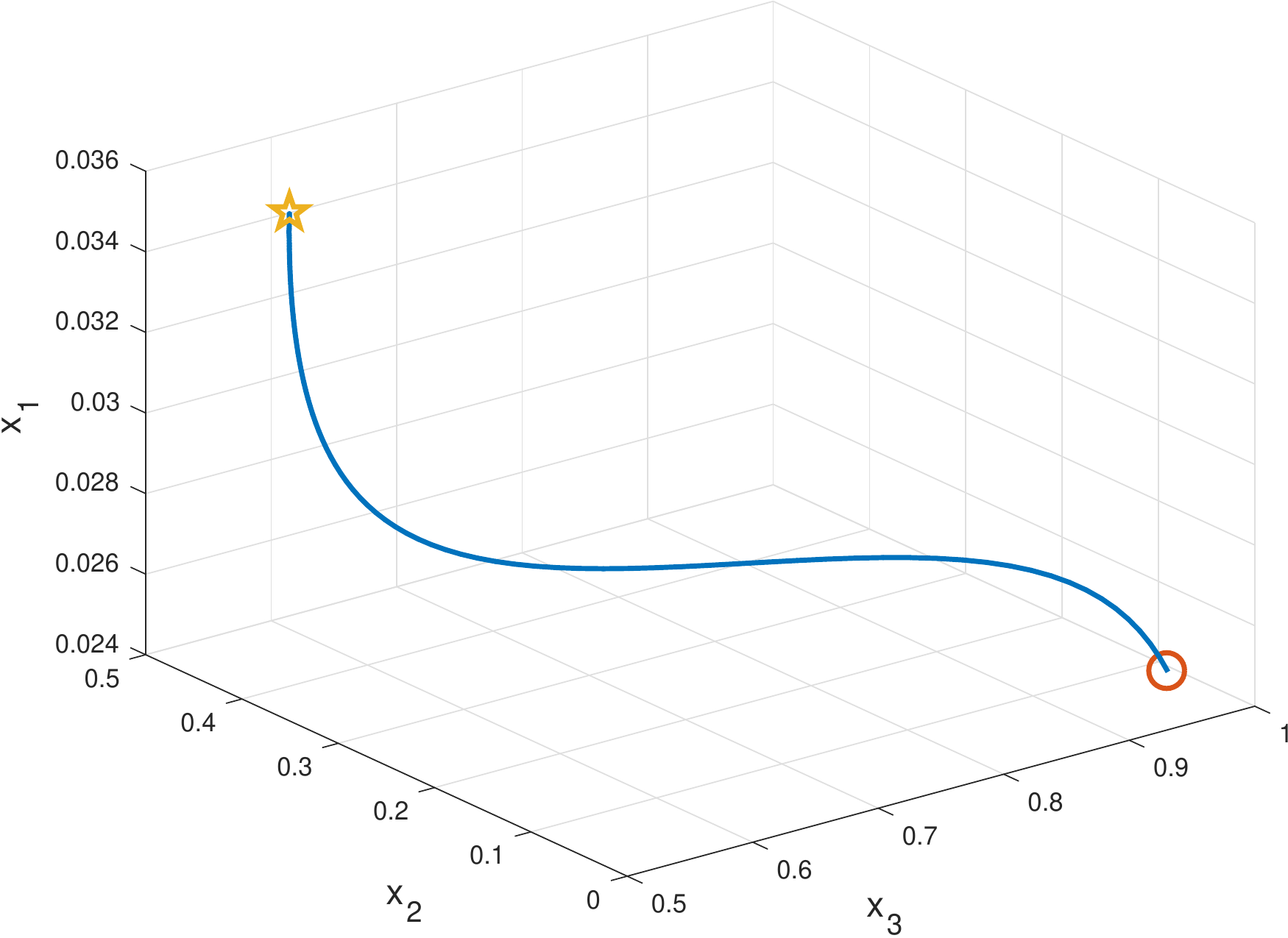}
\end{center}
\caption{A trajectory of the state of the 
replicator dynamic equation in \eqref{EQ:rde} with the initial state
of $(x_1, x_2, x_3) = (0.025, 0.025, 0.95)$ 
that is represented by $\circ$ marker, 
when $R = 1$, $c = 2$, $\Gamma = 4$ (or 6 dB),
$\bar \gamma = 10$ (or 10 dB), $\mu  = 0.2$.
The replicator dynamic converges to 
$(x_1^*, x_2^*, x_3^*) = (0.035, 0.415, 0.550)$,
which is represented by $\star$ marker.}
        \label{Fig:trj_3d}
\end{figure}

To find the ESS, we can use the replicator dynamic equation
in \eqref{EQ:rde} and an illustration of the trajectory of the state
is shown in Fig.~\ref{Fig:trj_3d}
when $(R, c) = (1, 2)$, $\Gamma = 6$ dB, $\bar \gamma = 10$ dB,
and $\mu = 0.2$.
By solving \eqref{EQ:cx1} and \eqref{EQ:R2C2},
we find that the ESS is given by
$$
(x_1^*, x_2^*, x_3^*) = (0.035, 0.415, 0.550),
$$
which can also be found by the replicator dynamic equation
after a sufficient number of iterations
as demonstrated in Fig.~\ref{Fig:trj_3d}.

\begin{figure}[thb]
\begin{center}
\includegraphics[width=\figwidth]{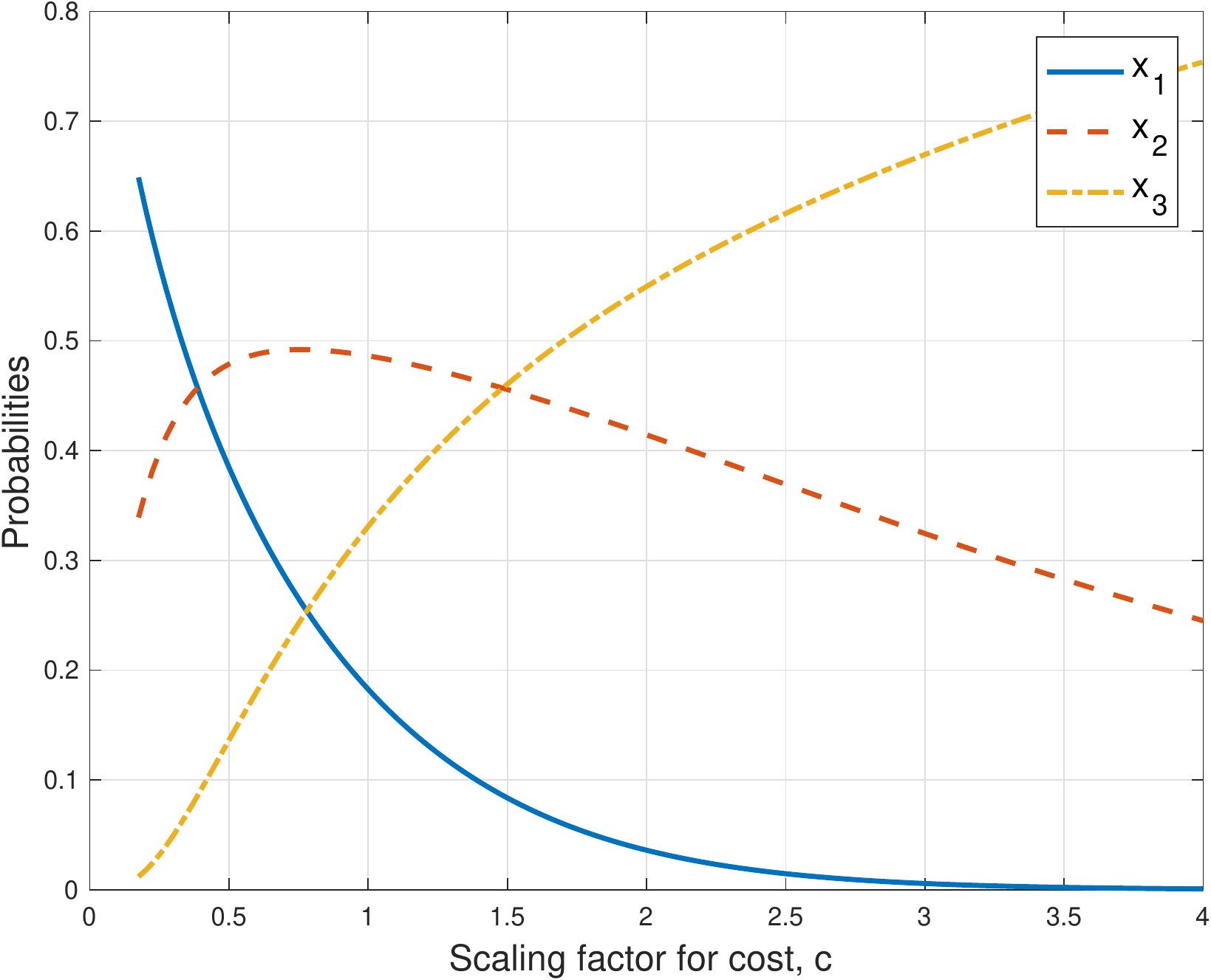} \\
(a) \\
\includegraphics[width=\figwidth]{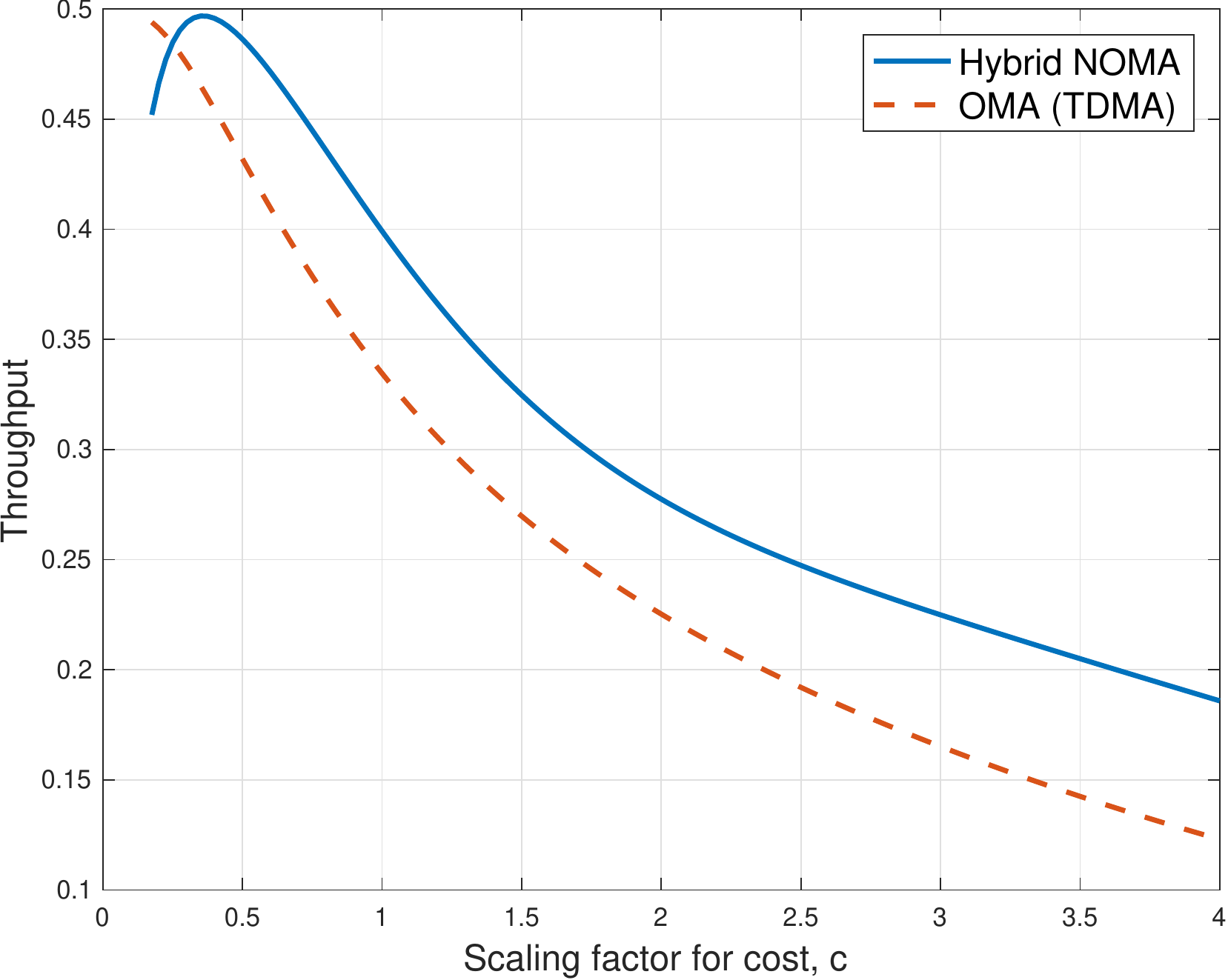} \\
(b) \\
\end{center}
\caption{Evolutionary game for hybrid uplink NOMA
with different values of the scaling factor for cost, $c$,
when $R = 1$, $\Gamma = 4$ (i.e., 6 dB), and $\bar \gamma = 10$ (i.e.,
10 dB): (a) ESS as a function of $c$; (b) throughput per user.}
        \label{Fig:trj1}
\end{figure}

Fig.~\ref{Fig:trj1} shows the results of
the evolutionary game for hybrid uplink NOMA
with different values of the scaling factor for cost, $c$,
when $R = 1$, $\Gamma = 6$ dB, and $\bar \gamma = 10$ dB. 
The ESS as a function of $c$ is shown in Fig.~\ref{Fig:trj1} (a),
where we can see that $x_1^*$ decreases with $c$,
which is expected by Property~\ref{L:5}.
That is, as the cost of transmission increases,
users are not encouraged to use action 1. 
It can be also observed that $x_1=x_2\approx0.46$ around $c=0.4$. This shows that the users choose action 1 or 2 equally likely so that a fair access can be achieved, while the maximum of throughput per user is obtained in Fig.~\ref{Fig:trj1} (b), 
where the throughput (per user) of hybrid uplink NOMA with
is compared to that  of OMA (i.e.,, TDMA).
The throughput of OMA is given by $\frac{1- x_3}{2}$. 
We can observe that for a large cost of transmission 
(i.e., a large $c$), $x_3$ becomes high. In this case,
the throughput of hybrid uplink NOMA is better 
than that of OMA as expected.
That is, with a large threshold $\tau$ for 
truncated channel inversion power control
(or a high $x_3$), it is better to share the channel 
with another user using NOMA to improve the throughput.
Furthermore, with a sufficiently high $x_3$,
as in \cite{ChoiJSAC}
\cite{Choi18}
\cite{Seo19}, more than two users
can be allocated to the same radio resource block.

\begin{figure}[thb]
\begin{center}
\includegraphics[width=\figwidth]{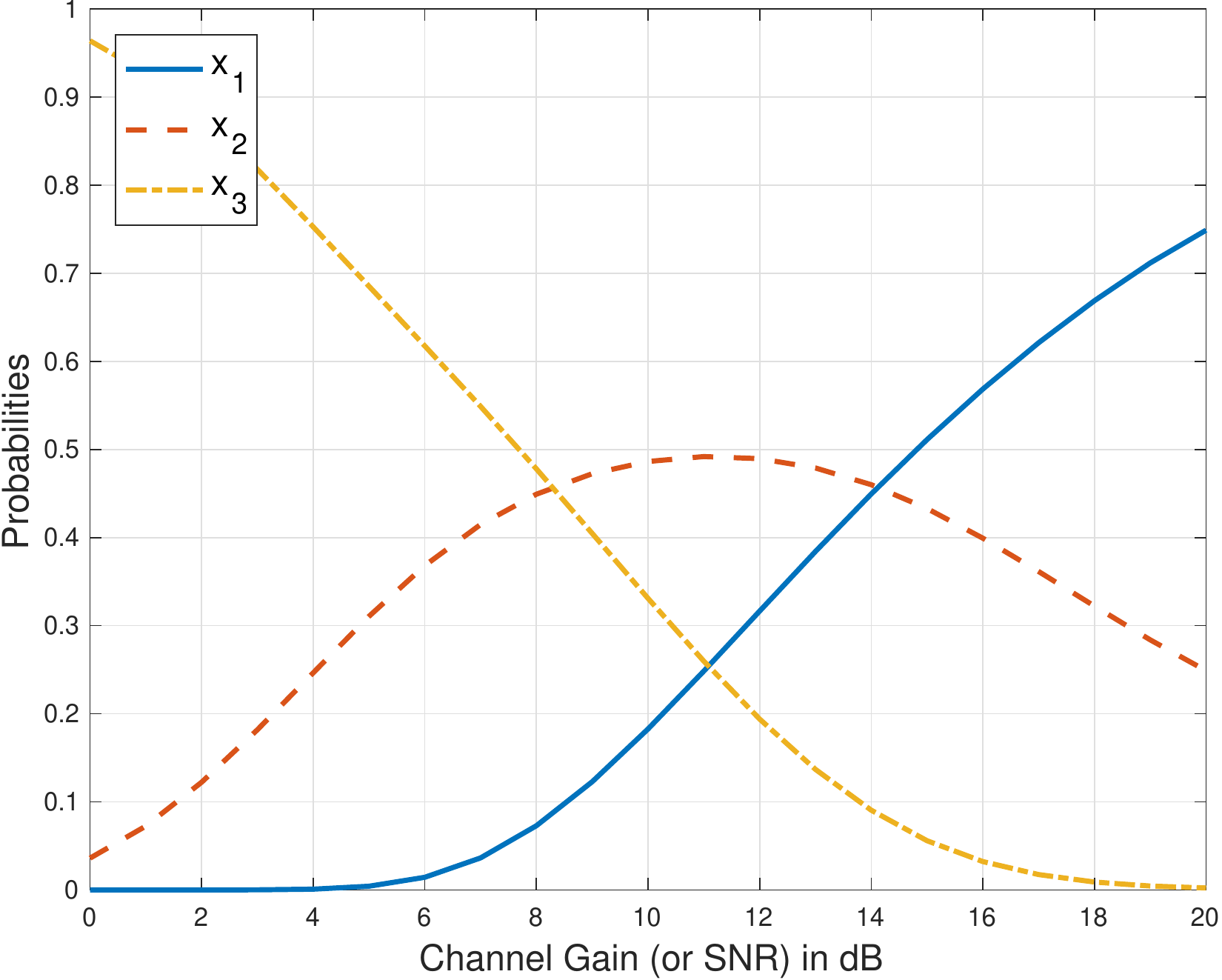} \\
(a) \\
\includegraphics[width=\figwidth]{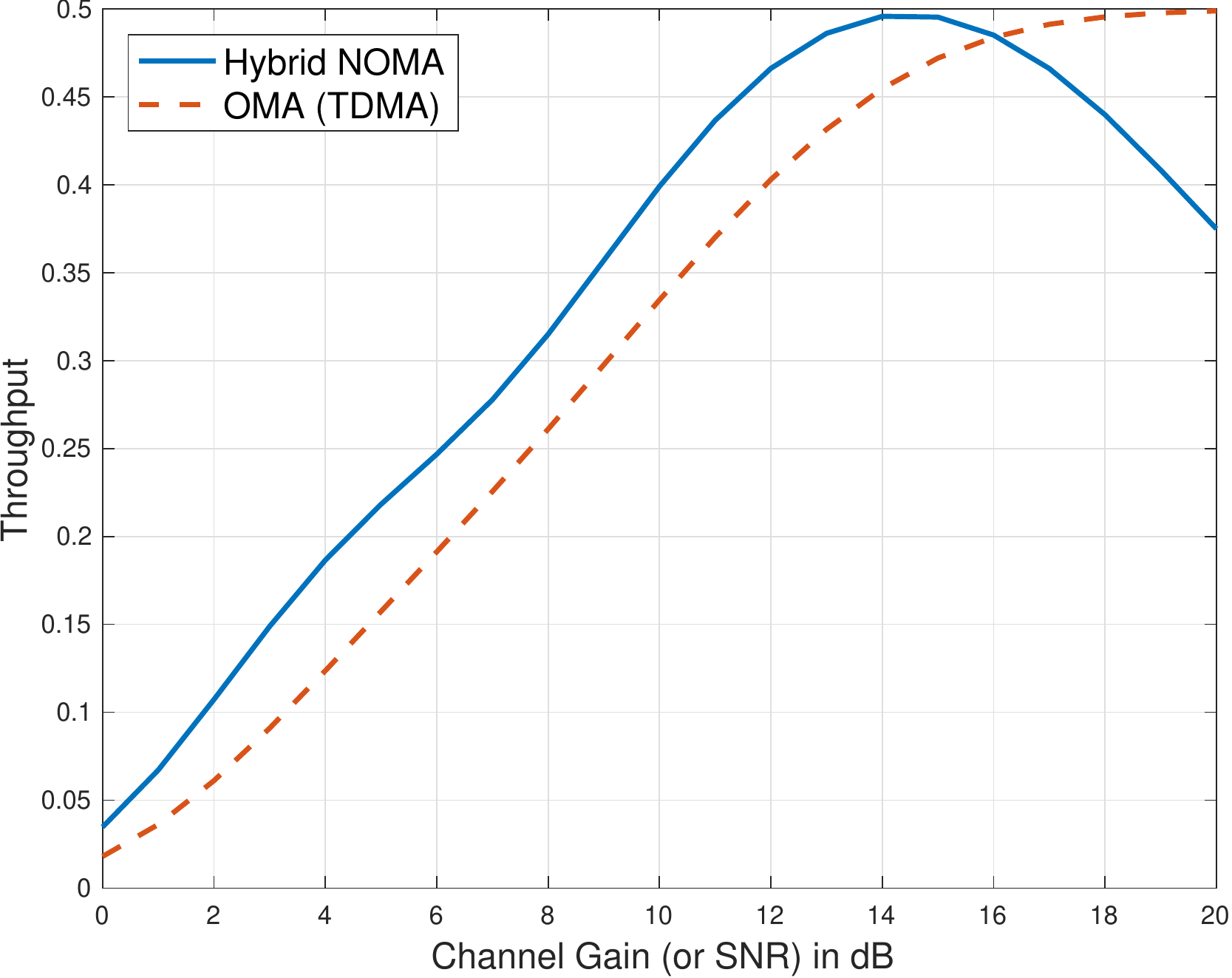} \\
(b) \\
\end{center}
\caption{Evolutionary game for hybrid uplink NOMA
with different values of average channel SNR, $\bar \gamma$
when $R = 1$, $c = 1$, and $\Gamma = 4$ (i.e., 6 dB):
(a) ESS as a function of $c$; (b) throughput per user.}
        \label{Fig:trj2}
\end{figure}

We show the results of
the evolutionary game for hybrid uplink NOMA
with different values of average channel SNR, $\bar \gamma$,
when $(R, c) = (1,1)$, $\Gamma = 6$ dB, and $\bar \gamma = 10$ dB
in Fig.~\ref{Fig:trj2}.
In Fig.~\ref{Fig:trj2} (a),
it is shown that
$x_1^*$ increases with $\bar \gamma$ as
expected by Property~\ref{L:5}. That is,
when $c$ is fixed, since the cost
decreases with $\bar \gamma$, users 
are more encouraged to employ action 1 for a higher $\bar \gamma$.
The throughput of hybrid uplink NOMA is shown in Fig.~\ref{Fig:trj2} (b),
where we can see that hybrid uplink NOMA
can have a higher throughput than OMA if $\bar \gamma$ is not too high.

\begin{figure}[thb]
\begin{center}
\includegraphics[width=\figwidth]{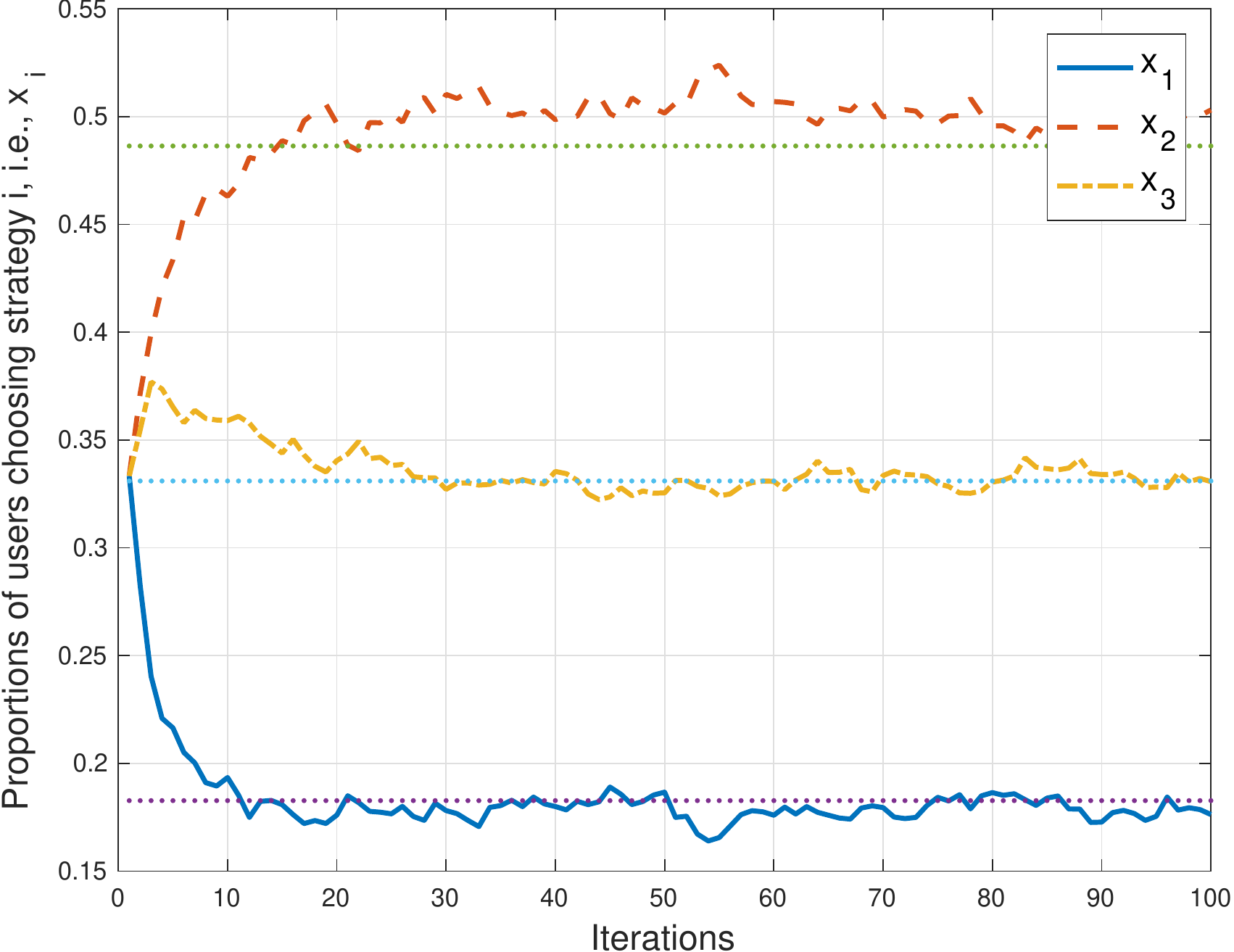}
\end{center}
\caption{A trajectory of $\bx$ obtained
by the replicator
dynamic equation in SU-BS
when $M = 300$, $\mu = 0.5$, $B = 40$,
$\Gamma = 6$ dB, $\bar \gamma = 10$ dB, and $(R, c) = (1,1)$.} 
        \label{Fig:Ugame1}
\end{figure}

As mentioned earlier, the ESS 
can be found using the replicator dynamic equation
with the time averages of the rewards and costs as their
estimates. To this end, in Subsection~\ref{SS:Impl}, 
we have discussed the state updating rules
at the BS and users, i.e., SU-BS and SU-U, respectively.
Fig.~\ref{Fig:Ugame1} 
shows the trajectory of $\bx$ 
obtained by the replicator
dynamic equation in SU-BS
when $M = 300$, $\mu = 0.5$, $B = 40$,
$\Gamma = 6$ dB, $\bar \gamma = 10$ dB, and $(R, c) = (1,1)$.
Note that the initial $\bx$ is set to $(1/3,1/3,1/3)$.
In Fig.~\ref{Fig:Ugame1},
the time for each iteration corresponds to one block interval
(i.e., the duration of $B = 40$ time slots).
We can observe that 
SU-BS 
(using the replicator dynamic equation with
the estimates of the average payoff 
that are obtained from time averages of the rewards and costs)
can provide a good estimate of the ESS 
that is $(x_1^*, x_2^*, x_3^*)
= (0.183, 0.486, 0.331)$.

\begin{figure}[thb]
\begin{center}
\includegraphics[width=\figwidth]{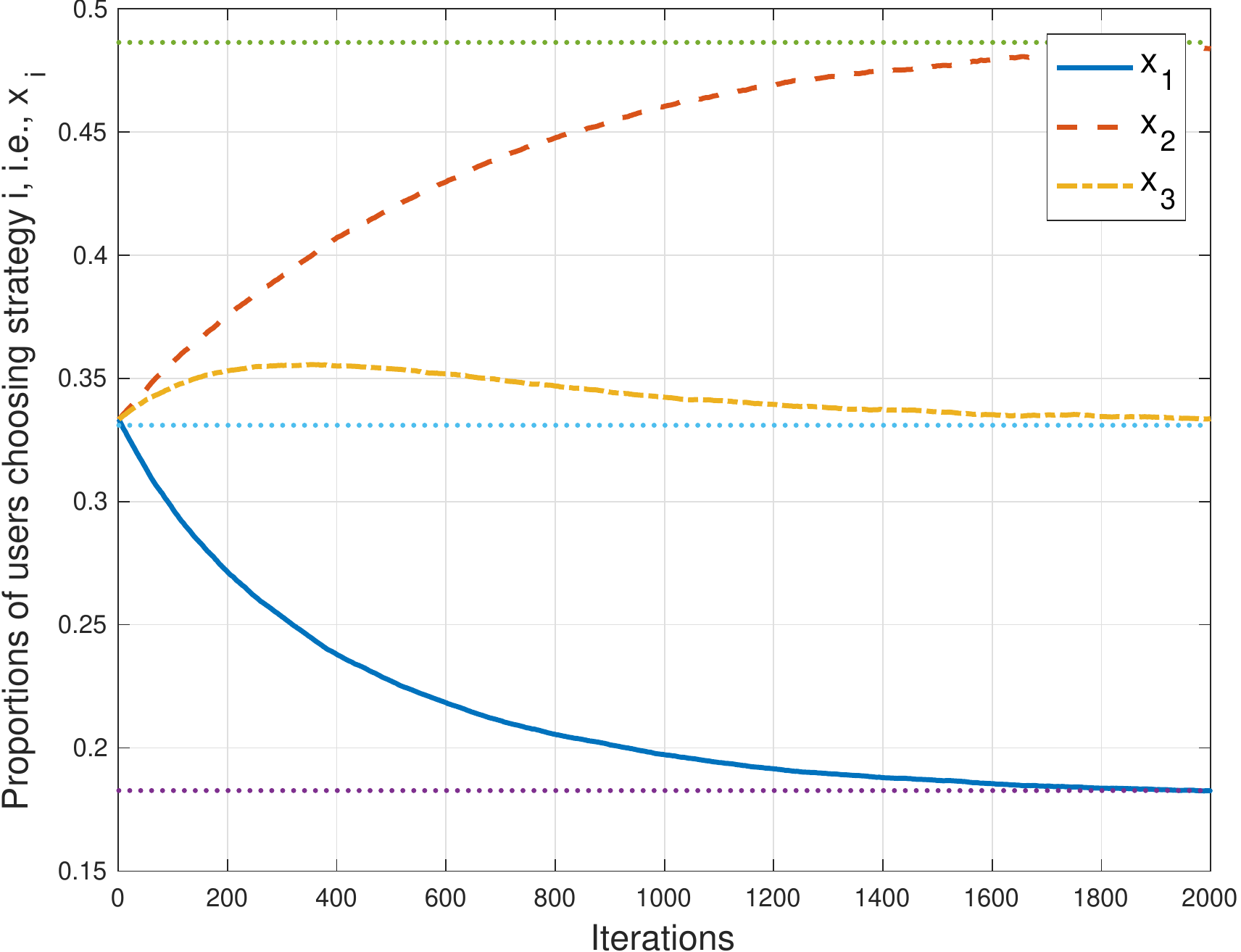}
\end{center}
\caption{A trajectory of $\bx$ 
obtained by the replicator
dynamic equation in SU-U
when $M = 300$, $\mu = 0.5$, $B = 40$,
$\Gamma = 6$ dB, $\bar \gamma = 10$ dB, and $(R, c) = (1,1)$.} 
        \label{Fig:Ugame2}
\end{figure}

Unlike SU-BS, SU-U is a distributed state updating rule
where each user updates the state 
and each user' state can be different from the others.
Thus, we show the average of $2 M$ users' states 
to show the trajectory of $\bx$ in Fig.~\ref{Fig:Ugame2}
where the state obtained by the replicator
dynamic equation in SU-U is shown  
when $M = 300$, $\mu = 0.5$, $B = 40$,
$\Gamma = 6$ dB, $\bar \gamma = 10$ dB, and $(R, c) = (1,1)$.
Compared with SU-BS in Fig.~\ref{Fig:Ugame2},
SU-U has a slow convergence
rate as it requires more iterations to converge to
the ESS.

\begin{figure}[thb]
\begin{center}
\includegraphics[width=\figwidth]{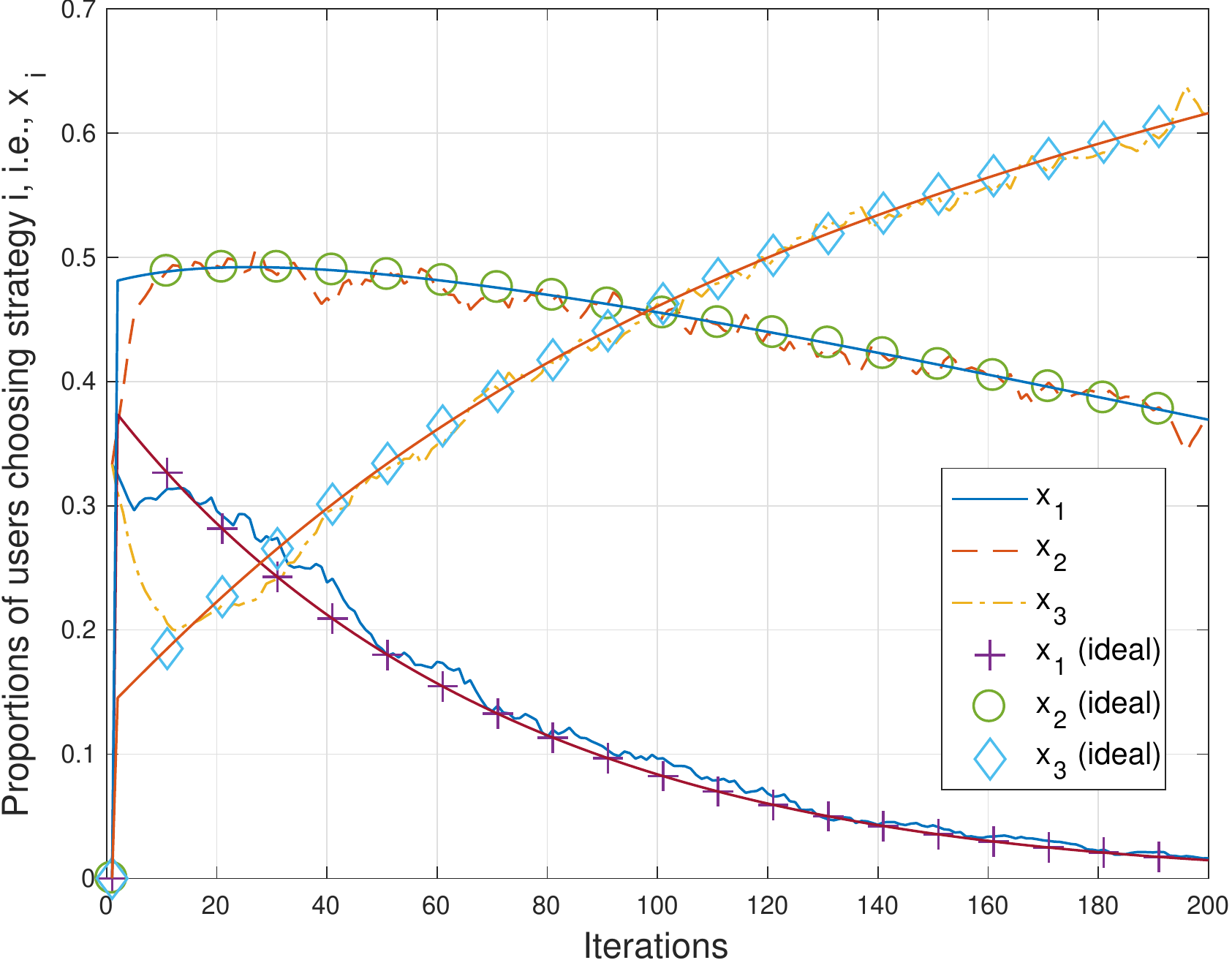}
\end{center}
\caption{A trajectory of $\bx$ obtained 
by the replicator
dynamic equation in SU-BS and the ESS of ideal case
(with the exact average payoffs)
with time-varying scaling factor for costs 
when $M = 300$, $\mu = 0.5$, $B = 40$,
$\Gamma = 6$ dB, $\bar \gamma = 10$ dB, and $R = 1$.} 
        \label{Fig:plt_UG1b}
\end{figure}

Since the replicator dynamic equation
can be seen as an adaptive updating rule for the state, it 
may be used when a parameter is varying over the time.
For example, the control of the scaling factor for
costs, $c$, might be necessary to improve
the throughput as shown in Fig.~\ref{Fig:trj1} (b).
In particular, the BS can broadcast a desirable 
value of $c$ to improve the overall performance.
To see how the state can be updated with the
replicator dynamic equation
in SU-B, we consider the following variation
of the scaling factor for costs for each block:
\be
c[b] = \frac{2 b}{200} + 0.5, \ b = 1, \ldots, 200,
	\label{EQ:cb1}
\ee
where $200$ is the number of blocks in a test.
In Fig.~\ref{Fig:plt_UG1b}, we show the trajectory of $\bx$ 
by the replicator dynamic equation in SU-BS
when $M = 300$, $\mu = 0.5$, $B = 40$,
$\Gamma = 6$ dB, $\bar \gamma = 10$ dB, and $R = 1$.
For comparisons, we also show the ESS 
with increasing $c[b]$
when the exact average payoffs are available.
It is shown that the trajectory of $\bx$ in SU-B
can closely follow the ESS after some iterations.

\begin{figure}[thb]
\begin{center}
\includegraphics[width=\figwidth]{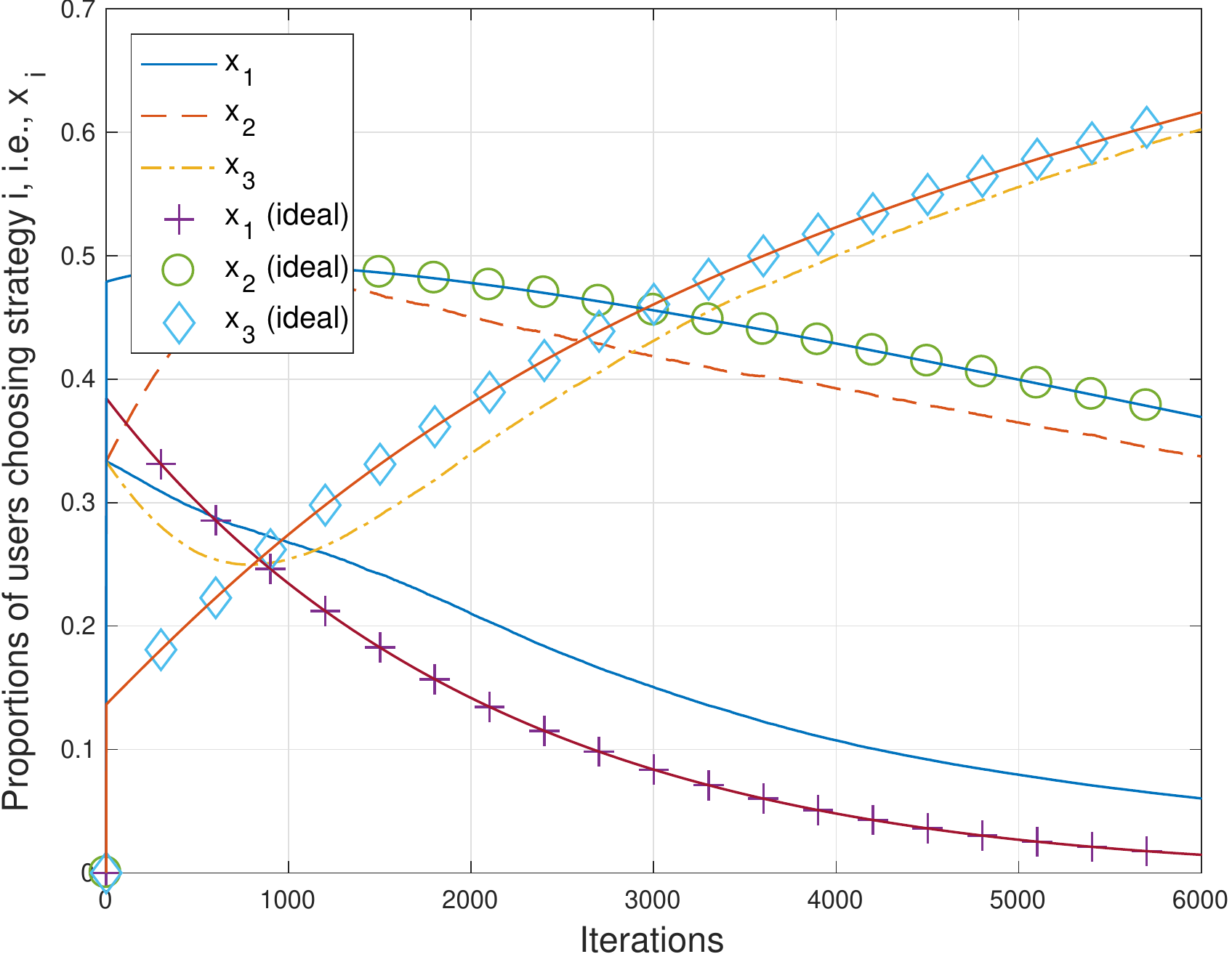}
\end{center}
\caption{A trajectory of $\bx$ obtained 
by the replicator
dynamic equation in SU-U and the ESS of ideal case
(with the exact average payoffs)
with time-varying scaling factor for costs 
when $M = 300$, $\mu = 0.5$, $B = 40$,
$\Gamma = 6$ dB, $\bar \gamma = 10$ dB, and $R = 1$.} 
        \label{Fig:plt_UG2b}
\end{figure}

To see the trajectory of $\bx$ in SU-U for varying 
$c$, we consider the variation
of the scaling factor for costs for each block as follows:
\be
c[b] = \frac{2 b}{6000} + 0.5, \ b = 1, \ldots, 6000,
	\label{EQ:cb2}
\ee
where $6000$ is the number of blocks in a test.
Note that the variation of $c$ in \eqref{EQ:cb2}
is slower than that in  \eqref{EQ:cb1} by  a factor of
$\frac{6000}{200} = 30$.
In Fig.~\ref{Fig:plt_UG2b}, we show the trajectory of $\bx$ 
by the replicator dynamic equation in SU-U 
when $M = 300$, $\mu = 0.5$, $B = 40$,
$\Gamma = 6$ dB, $\bar \gamma = 10$ dB, and $R = 1$.
When we compare the trajectory 
in SU-U with the ESS,
it is clear that there is a time lag\footnote{A large step-size $\mu$
can be used for a better tracking performance. However,
a large step-size $\mu$ leads to instability
although it is not shown in the paper.  
As a result, the selection of 
the step-size has to be carefully considered, which 
is beyond the scope of the paper and might be a further research
topic.}.

From Figs.~\ref{Fig:plt_UG1b} and~\ref{Fig:plt_UG2b},
we can see that SU-BS can update the state faster than SU-U
in accordance with the variation of $c$.
Thus, SU-BS is preferable to SU-U when certain key parameters 
(e.g., $c$) are controlled by the BS 
for better performance at the cost of high signaling overhead.


\section{Concluding Remarks}	\label{S:Conc}

We proposed a hybrid uplink NOMA scheme
to support more users using power-domain NOMA.
In order to avoid high signaling overhead for the power allocation 
that is usually required for power-domain NOMA,
truncated channel inversion power control
at users was considered.
The proposed hybrid uplink NOMA scheme
was able to exploit fading in such a way that
when one user does not transmit signals due to severe fading,
another user can access the radio resource block.
To decide the threshold values for 
the truncated channel inversion power control in hybrid uplink NOMA,
an evolutionary game was formulated and its solution
(i.e., ESS) was characterized. 
We also showed that the replicator dynamic equation
can be used to find the ESS and discussed two implementation approaches
to update the state with outcomes 
about the success of transmissions and realizations of fading channels.

Note that in this paper, we focused on
the power control strategy for the hybrid uplink NOMA scheme
based on power-domain NOMA. 
As in \cite{Chen18}, there are other uplink NOMA schemes (e.g.,
sparse code multiple access (SCMA)).
As a further research topic, the sparse code control
(which might be equivalent to the power control)
can be studied for SCMA from a point of view of evolutionary
game theory. In addition, although
we only consider two users 
per radio resource block in this paper,
a generalization with more than two users per radio
resource block is possible with more power levels. 
This generalization with a combination of
power-domain NOMA and SCMA
(to keep the maximum transmit power limited when
there are a number of users per radio resource block)
might be an interesting topic to be studied in the future.

\appendices

\section{Proof of Property~\ref{L:1}}	\label{A:1}

For convenience, denote by $\uP_\bx (i)$
the probability that user 1 succeeds with action $i 
\in \{1,2\}$ when the state 
(or the mixed strategy) of user 2 is $\bx$.
Then, it can be shown that
\begin{align}
\uP_\bx (1) = x_2 + x_3 \ \mbox{and} \
\uP_\bx (2) = x_1 + x_3.
	\label{EQ:PPx}
\end{align}
In \eqref{EQ:PPx}, we consider SIC to find $\uP_\bx (2)$.
That is, although user 2 chooses action 1, 
user 1 can still succeed with action 2 using SIC.
Since
$$
\uE_\bx [\indicator({\rm succeed \ with \ action}\ i)]
= \uP_\bx (i), \ i \in \{1,2\},
$$
we can show that
\begin{align}
u_1 (i, \bx) = R_i (\bx) - \bar C_i, \ i \in \{1,2\},
	\label{EQ:uix}
\end{align}
where
\begin{align}
R_i (\bx) = 
\left\{
\begin{array}{ll}
R (x_2 + x_3), & \mbox{if $i = 1$} \cr
R (x_1 + x_3), & \mbox{if $i = 2$.} \cr
\end{array}
\right.
	\label{EQ:Rx23}
\end{align}
Substituting \eqref{EQ:uix} into 
the first equation in \eqref{EQ:uxx}, we can have
the second equation in \eqref{EQ:uxx} after some manipulations.
This completes the proof.

\section{Proof of Property~\ref{L:2}}	\label{A:2}

With a slight abuse of notation,
let $u_k (i,j)$ denote the payoff of user $k$ when
user 1 and user 2 choose pure strategies, $i$ and $j$.
In addition, if 
user 1 chooses a mixed strategy $\bx$ and 
user 2 chooses a mixed strategy $\bx^\prime$,
the payoff of user $k$ is denoted by $u_k (\bx, \bx^\prime)$.
If user 2 chooses a mixed strategy $\bx$ and user 1 chooses
a pure strategy $i$, then the payoff of user 1 is denoted by
$u_1 (i, \bx)$, which is identical to that in 
\eqref{EQ:u_1}.
By the definition of a mixed strategy NE \cite{Maschler13}
we have
\be
u_1 (i, \bx^*) \le u_1 (\bx^*, \bx^*), \ i \in \{1,2,3\},
	\label{EQ:uu1}
\ee
if $\bx^*$ is a mixed strategy NE.
Note that since the game is symmetric,
\eqref{EQ:uu1} is equivalent to
$u_2 (\bx^*, i) \le u_2 (\bx^*, \bx^*), \ i \in \{1,2,3\}$.
From \eqref{EQ:uxx}, it can be readily shown that
\be
u_1(\bx^*, \bx) = u(\bx^*, \bx^*).
	\label{EQ:u1u}
\ee
Substituting \eqref{EQ:u1u} into \eqref{EQ:uu1}, we can have
\eqref{EQ:NE}, which completes the proof.

\section{Proof of Property~\ref{L:3}}	\label{A:3}

In case of A,
according to the indifference principle \cite{Maschler13},
we have
\be
u_1(1, \bx^*) = u_1(2, \bx^*) = u_1(3, \bx^*).
	\label{EQ:u3}
\ee
Since $x_1^* + x_2^* + x_3^* = 1$, 
from \eqref{EQ:u3}, we have
$$
R x_1^*- R + C_1 = R x_2^*- R + C_2 = 0.
$$
This leads to \eqref{EQ:L_A} as $C_1 < R$ and $C_1 + C_2 > R$. 

In case of B, we can see that $x_1^*$ has to be zero,
while $x_2^*, x_3^* > 0$.
According to the indifference principle \cite{Maschler13},
we need to have $u_1(2,\bx^*) = u_1(3, \bx^*)$,
which results in \eqref{EQ:L_B}.

In case of C, 
since costs $C_1$ and $C_2$ are higher than reward $R$,
$x_1^*$ and $x_2^*$ become 0 and $x_3^* = 1$, which means
that no transmission becomes NE. 

In case of D, we can see that
$u_1(1, \bx) = u_1(2, \bx)$ with $x_3^* = 0$, which leads
to \eqref{EQ:L_D} and completes the proof.

\section{Proof of Property~\ref{L:4}}	\label{A:4}

From \eqref{EQ:hk},
the probability density function (pdf) of $\gamma_k$ is given by
$f_\gamma (\gamma_k) = \frac{1}{\bar \gamma}
e^{-\frac{\gamma_k}{\bar \gamma} }$.
Using this, we can show that
\begin{align}
\bar C_1 & = 
c \rho_1 
\uE \left[ \frac{1}{\gamma_k} \,|\, \gamma_k \in \cG_1\right] \cr
& = c \rho_1 
\int_0^\infty \frac{1}{\gamma} f_{\gamma} 
(\gamma \,|\, \gamma \in \cG_1) d \gamma \cr
& = \frac{c \rho_1}{x_1}
\int_{\tau_{\rm pn}}^\infty \frac{1}{\gamma} f_\gamma (\gamma) d \gamma 
= \frac{c \rho_1}{\bar \gamma x_1}
E_1 \left( \frac{\tau_{\rm pn}}{\bar \gamma} \right).
	\label{EQ:aC1}
\end{align}
Similarly, we can also derive that
\be
\bar C_2 = 
\frac{c \rho_2}{\bar \gamma x_2}
\left( E_1 \left( \frac{\tau}{\bar \gamma} \right)
-E_1 \left( \frac{\tau_{\rm pn}}{\bar \gamma} \right) \right).
	\label{EQ:aC2}
\ee

Furthermore, 
from \eqref{EQ:hk} and \eqref{EQ:pxi}, we can readily show that
\begin{align}
\tau & = \bar \gamma \ln \frac{1}{1 - x_3} 
= \bar \gamma \ln \frac{1}{x_1 + x_2} \cr
\tau_{\rm pn} & = \bar \gamma \ln \frac{1}{x_1}.
	\label{EQ:tts}
\end{align}
Substituting \eqref{EQ:tts} into \eqref{EQ:aC1} and \eqref{EQ:aC2},
we have the costs in \eqref{EQ:CCs}, which completes the proof. 

\section{Proof of Property~\ref{L:5}}	\label{A:5}

Since $x_3 > 0$, based on the indifference principle,
we need to have
$u_1 (1, \bx) = 0$
or from \eqref{EQ:Rx23} and \eqref{EQ:CCs},
$$
R (x_2 + x_3) - \bar C_1 (\bx)
= R(1-x_1) - \bar C_1 (\bx) = 0,
$$
which leads to \eqref{EQ:cx1}.
Consider
$V(x) = \frac{ E_1 \left( \ln \frac{1}{x} \right)}{x}$, $x \in (0,1)$.
It can be shown that
\begin{align}
\lim_{x \to 0} V(x) =  \lim_{x \to 0} 
\frac{\frac{1}{x} E_0 (-\ln x) }{1} = 0,
\end{align}
where the limit is due to L'Hospital's rule,
the fact that $\frac{d}{dx} E_n (x) = - E_{n-1}(x)$ (for $n = 0,1\ldots$),
and $E_0(x) = -\frac{e^{-x}}{x}$.
In addition, let
$\tilde V = V(\tilde x_3)$.
It can be shown that
\begin{align}
\frac{d V(x)}{dx} & = - \frac{1}{x \ln x}
- \frac{E_1 (-\ln x)}{x^2} \cr
& = \frac{1}{x} \left( - \frac{1}{\ln x} - \frac{E_1 (-\ln x)}{x} \right) \cr
& \ge 
\frac{1}{x} \left( - \frac{1}{\ln x} - \ln \left( 1 -
\frac{1}{\ln x}
\right)  \right)  > 0,
\end{align}
where the first inequality is due to
$E_1 (x) \le e^{-x} \ln \left(1 + \frac{1}{x} \right)$, $x \ge 0$, 
and the second inequality is due to $z > \ln (1+z)$, $z > 0$,
with $z = -\frac{1}{\ln x} > 0$.
Since $C_1 (\bx) = \frac{c \rho_1}{\bar \gamma} V(x)$,
$C_1(\bx)$ is an increasing function
of $x_1$ and $\lim_{x_1 \to 0} C_1(\bx) = 0$ and 
$$
C_1(\bx) \bigl|_{x_1 = \tilde x_3} =
\frac{c \rho_1}{\bar \gamma}  \tilde V,
$$
while $R(1-x_1)$ is a decreasing function of $x_1$.
As a result, \eqref{EQ:cx1}
has a unique solution
if
$\frac{c \rho_1}{\bar \gamma}  \tilde V > R x_3$,
which is equivalent to \eqref{EQ:x3_con}.

Furthermore, since 
$C_1 (\bx)$ increases 
with $c$ and decreases with $\bar \gamma$,
we can also see that $x_1^*$ decreases with $c$ 
and increases with $\bar \gamma$.

\bibliographystyle{ieeetr}
\bibliography{noma}

\begin{thebibliography}{10}

\bibitem{Dai15}
L.~Dai, B.~Wang, Y.~Yuan, S.~Han, C.~I, and Z.~Wang, ``Non-orthogonal multiple
  access for {5G}: solutions, challenges, opportunities, and future research
  trends,'' {\em IEEE Communications Magazine}, vol.~53, pp.~74--71, September
  2015.

\bibitem{Ding_CM}
Z.~Ding, Y.~Liu, J.~Choi, M.~Elkashlan, C.~L. I, and H.~V. Poor, ``Application
  of non-orthogonal multiple access in {LTE} and {5G} networks,'' {\em IEEE
  Communications Magazine}, vol.~55, pp.~185--191, February 2017.

\bibitem{Choi17_ISWCS}
J.~Choi, ``{NOMA}: Principles and recent results,'' in {\em 2017 International
  Symposium on Wireless Communication Systems (ISWCS)}, pp.~349--354, Aug 2017.

\bibitem{Dai18}
L.~{Dai}, B.~{Wang}, Z.~{Ding}, Z.~{Wang}, S.~{Chen}, and L.~{Hanzo}, ``A
  survey of non-orthogonal multiple access for 5g,'' {\em IEEE Communications
  Surveys Tutorials}, vol.~20, pp.~2294--2323, thirdquarter 2018.

\bibitem{Saito13}
Y.~Saito, Y.~Kishiyama, A.~Benjebbour, T.~Nakamura, A.~Li, and K.~Higuchi,
  ``Non-orthogonal multiple access ({NOMA}) for cellular future radio access,''
  in {\em Vehicular Technology Conference (VTC Spring), 2013 IEEE 77th},
  pp.~1--5, June 2013.

\bibitem{Kim13}
B.~Kim, S.~Lim, H.~Kim, S.~Suh, J.~Kwun, S.~Choi, C.~Lee, S.~Lee, and D.~Hong,
  ``Non-orthogonal multiple access in a downlink multiuser beamforming
  system,'' in {\em MILCOM 2013 - 2013 IEEE Military Communications
  Conference}, pp.~1278--1283, Nov 2013.

\bibitem{Sun18}
Y.~{Sun}, Z.~{Ding}, and X.~{Dai}, ``On the performance of downlink {NOMA} in
  multi-cell mmwave networks,'' {\em IEEE Communications Letters}, vol.~22,
  pp.~2366--2369, Nov 2018.

\bibitem{Xiao18}
Z.~{Xiao}, L.~{Zhu}, J.~{Choi}, P.~{Xia}, and X.~{Xia}, ``Joint power
  allocation and beamforming for non-orthogonal multiple access ({NOMA}) in
  {5G} millimeter wave communications,'' {\em IEEE Trans. Wireless
  Communications}, vol.~17, pp.~2961--2974, May 2018.

\bibitem{Choi14}
J.~Choi, ``Non-orthogonal multiple access in downlink coordinated two-point
  systems,'' {\em IEEE Commun. Letters}, vol.~18, pp.~313--316, Feb. 2014.

\bibitem{Shin17}
W.~Shin, M.~Vaezi, B.~Lee, D.~J. Love, J.~Lee, and H.~V. Poor, ``Coordinated
  beamforming for multi-cell {MIMO-NOMA},'' {\em IEEE Communications Letters},
  vol.~21, pp.~84--87, Jan 2017.

\bibitem{Sun18a}
Y.~{Sun}, Z.~{Ding}, X.~{Dai}, and G.~K. {Karagiannidis}, ``A feasibility study
  on network {NOMA},'' {\em IEEE Trans. Communications}, vol.~66,
  pp.~4303--4317, Sep. 2018.

\bibitem{Sun19}
Y.~{Sun}, Z.~{Ding}, X.~{Dai}, and O.~A. {Dobre}, ``On the performance of
  network {NOMA} in uplink {CoMP} systems: A stochastic geometry approach,''
  {\em IEEE Trans. Communications}, pp.~1--1, 2019.

\bibitem{Imari14}
M.~Al-Imari, P.~Xiao, M.~A. Imran, and R.~Tafazolli, ``Uplink non-orthogonal
  multiple access for {5G} wireless networks,'' in {\em 2014 11th International
  Symposium on Wireless Communications Systems (ISWCS)}, pp.~781--785, Aug
  2014.

\bibitem{Zhang16}
N.~{Zhang}, J.~{Wang}, G.~{Kang}, and Y.~{Liu}, ``Uplink nonorthogonal multiple
  access in {5G} systems,'' {\em IEEE Communications Letters}, vol.~20,
  pp.~458--461, March 2016.

\bibitem{Choi17_CSI}
J.~Choi, ``Joint rate and power allocation for {NOMA} with statistical {CSI},''
  {\em IEEE Trans. Communications}, vol.~65, pp.~4519--4528, Oct 2017.

\bibitem{Liu18}
Y.~{Liu}, M.~{Derakhshani}, and S.~{Lambotharan}, ``Outage analysis and power
  allocation in uplink non-orthogonal multiple access systems,'' {\em IEEE
  Communications Letters}, vol.~22, pp.~336--339, Feb 2018.

\bibitem{ChoiJSAC}
J.~Choi, ``{NOMA}-based random access with multichannel {ALOHA},'' {\em IEEE J.
  Selected Areas in Communications}, vol.~35, pp.~2736--2743, Dec 2017.

\bibitem{Choi18_L}
J.~Choi, ``Layered non-orthogonal random access with {SIC} and transmit
  diversity for reliable transmissions,'' {\em IEEE Trans. Communications},
  vol.~66, pp.~1262--1272, March 2018.

\bibitem{Fudenberg}
D.~Fudenberg and J.~Tirole, {\em Game Theory}.
\newblock Cambridge, MA: MIT Press, 1991.

\bibitem{Maschler13}
M.~Maschler, S.~Zamir, and E.~Solan, {\em Game Theory}.
\newblock Cambridge University Press, 2013.

\bibitem{Choi_18G}
J.~Choi, ``Multichannel {NOMA-ALOHA} game with fading,'' {\em IEEE Trans.
  Communications}, vol.~66, pp.~4997--5007, Oct 2018.

\bibitem{Seo18}
J.~{Seo} and H.~{Jin}, ``Two-user {NOMA} uplink random access games,'' {\em
  IEEE Communications Letters}, vol.~22, pp.~2246--2249, Nov 2018.

\bibitem{Seo19}
J.~{Seo}, T.~{Kwon}, and J.~{Choi}, ``Evolutionary game approach to uplink
  {NOMA} random access systems,'' {\em IEEE Communications Letters}, pp.~1--1,
  2019.

\bibitem{Ding19}
Z.~{Ding}, R.~{Schober}, P.~{Fan}, and H.~V. {Poor}, ``Simple semi-grant-free
  transmission strategies assisted by non-orthogonal multiple access,'' {\em
  IEEE Trans. Communications}, pp.~1--1, 2019.

\bibitem{Goldsmith97a}
A.~Goldsmith and P.~Varaiya, ``Capacity of fading channels with channel side
  information,'' {\em IEEE Trans. Inform. Theory}, vol.~43, pp.~1986--1992,
  Nov. 1997.

\bibitem{TseBook05}
D.~Tse and P.~Viswanath, {\em Fundamentals of Wireless Communication}.
\newblock Cambridge University Press, 2005.

\bibitem{Choi_18C}
J.~Choi, ``On power and rate allocation for coded uplink {NOMA} in a
  multicarrier system,'' {\em IEEE Trans. Communications}, vol.~66,
  pp.~2762--2772, June 2018.

\bibitem{Weibull95}
J.~Weibull, {\em Evolutionary Game Theory}.
\newblock Evolutionary Game Theory, MIT Press, 1995.

\bibitem{Timotheou15}
S.~Timotheou and I.~Krikidis, ``Fairness for non-orthogonal multiple access in
  {5G} systems,'' {\em IEEE Signal Process. Letters}, vol.~22, pp.~1647--1651,
  Oct 2015.

\bibitem{Choi16_F}
J.~Choi, ``Power allocation for max-sum rate and max-min rate proportional
  fairness in {NOMA},'' {\em IEEE Commun. Letters}, vol.~20, pp.~2055--2058,
  Oct 2016.

\bibitem{Choi18}
J.~{Choi}, ``Multichannel {NOMA-ALOHA} game with fading,'' {\em IEEE Trans.
  Communications}, vol.~66, pp.~4997--5007, Oct 2018.

\bibitem{Chen18}
Y.~{Chen}, A.~{Bayesteh}, Y.~{Wu}, B.~{Ren}, S.~{Kang}, S.~{Sun}, Q.~{Xiong},
  C.~{Qian}, B.~{Yu}, Z.~{Ding}, S.~{Wang}, S.~{Han}, X.~{Hou}, H.~{Lin},
  R.~{Visoz}, and R.~{Razavi}, ``Toward the standardization of non-orthogonal
  multiple access for next generation wireless networks,'' {\em IEEE
  Communications Magazine}, vol.~56, pp.~19--27, March 2018.

\end{thebibliography}
\end{document}